\documentclass[journal]{IEEEtran}
\usepackage{cite}
\usepackage{glossaries}
\usepackage{graphicx}
\usepackage{commath}
\usepackage{color}
\usepackage{amssymb}
\usepackage{mathtools}
\usepackage{empheq}
\usepackage{algorithm}
\usepackage{subcaption}
\usepackage{hyperref}

\newacronym{FNR}{FNR}{Luxembourg National Research Fund}
\newacronym{SNR}{SNR}{Signal to Noise Ratio}
\newacronym{INR}{INR}{Interfernece to Noise Ratio}
\newacronym{SINR}{SINR}{Signal to Interference plus Noise Ratio}
\newacronym{AF}{AF}{Ambiguity Function}
\newacronym{MIMO}{MIMO}{Multiple-Input Multiple-Output}
\newacronym{SISO}{SISO}{Single-Input Single-Output}
\newacronym{SIMO}{SIMO}{Single-Input Multiple-Output}
\newacronym{CD}{CD}{Coordinate Descent}
\newacronym{BCD}{BCD}{Block Coordinate Descent}
\newacronym{GD}{GD}{Gradient Descent}
\newacronym{MM}{MM}{Majorization-Minimization}
\newacronym{FMCW}{FMCW}{Frequency Modulated Continuous Wave}
\newacronym{PMCW}{PMCW}{Phase Modulated Continuous Wave}
\newacronym{DFT}{DFT}{Discrete Fourier Transform}
\newacronym{FFT}{FFT}{Fast Fourier Transform}
\newacronym{MVDR}{MVDR}{Minimum Variance Distortionless Response}
\newacronym{MBI}{MBI}{Maximum Block Improvement}
\newacronym{RFPA}{RFPA}{Radio Frequency Power Amplifier}
\newacronym{BPSK}{BPSK}{Binary Phase Shift Keying}
\newacronym{QPSK}{QPSK}{Quadrature Phase Shift Keying}
\newacronym{ULA}{ULA}{Uniform Linear Array}
\newacronym{DOF}{DOF}{Degrees of Freedom}
\newacronym{PSK}{PSK}{Phase Shift Keying}
\newacronym{PSL}{PSL}{Peak Sidelobe Level}
\newacronym{PSLR}{PSLR}{Peak Sidelobe Level Ratio}
\newacronym{ISL}{ISL}{Integrated Sidelobe Level}
\newacronym{ISLR}{ISLR}{Integrated Sidelobe Level Ratio}
\newacronym{LFM}{LFM}{Linear Frequency Modulation}
\newacronym{CPI}{CPI}{Coherent Pulse Interval}
\newacronym{RCS}{RCS}{Radar Cross Section}
\newacronym{CNR}{CNR}{Clutter to Noise Ratio}
\newacronym{MTI}{MTI}{Moving Target Indicator}
\newacronym{ROC}{ROC}{Receiver Operating Characteristic}
\newacronym{MPSK}{MPSK}{$M$-ary Phase Shift Keying}
\newacronym{PAR}{PAR}{Peak-to-Average Ratio}
\newacronym{GFP}{GFP}{Generalized Fractional Programming}
\newacronym{PRI}{PRI}{Pulse Repetition Interval}
\newacronym{PRF}{PRF}{Pulse Repetition Frequency}
\newacronym{MMM}{MM}{Majorization Minimization or Minorization Maximization}
\newacronym{QCQP}{QCQP}{Quadratic Constraint Quadratic Programming}
\newacronym{SDP}{SDP}{Semi-definite Programming}
\newacronym{CCL}{CCL}{Cross-Correlation Level}
\newacronym{TDMA}{TDMA}{Time-Division Multiple Access}
\newacronym{FDMA}{FDMA}{Frequency-Division Multiple Access}
\newacronym{CDMA}{CDMA}{Code-Division Multiple Access}
\newacronym{DDMA}{DDMA}{Doppler-Division Multiple Access}
\newacronym{TDM}{TDM}{Time Division Multiplexing}
\newacronym{FDM}{FDM}{Frequency Division Multiplexing}
\newacronym{CDM}{CDM}{Code Division Multiplexing}
\newacronym{DDM}{DDM}{Doppler Division Multiplexing}
\newacronym{SDR}{SDR}{Semi-definite Relaxation}
\newacronym{QSDR}{QSDR}{Quantized Semi-definite Relaxation}
\newacronym{CA}{CA}{Cyclic Algorithm}
\newacronym{ADMM}{ADMM}{Alternating Direction Method of Multipliers}
\newacronym{PDR}{PDR}{Projection, Descent, and Retraction}
\newacronym{SQP}{SQP}{Semidefinite Quadratic Programming}
\newacronym{CM}{CM}{Constant Modulus}
\newacronym{MPS}{MPS}{Minimum Peak Sidelobe}
\newacronym{BiST}{BiST}{Binary Sequences seTs}
\newacronym{ESA}{ESA}{Effective Simulated Annealing}
\newacronym{BSUM}{BSUM}{Block Successive Upper Bound Minimization}
\newacronym{CS}{CS}{Compressive Sensing}
\newacronym{CAN}{CAN}{Cyclic Algorithm-New}
\newacronym{CW}{CW}{Continuous Wave}
\newacronym{WeBEST}{WeBEST}{Weighted BSUM sEquence SeT}
\newacronym{MISL}{MISL}{Monotonic minimizer for Integrated Sidelobe Level}
\newacronym{DP}{DP}{Discrete Phase}
\newacronym{CP}{CP}{Continuous Phase}
\newacronym{SAR}{SAR}{Synthetic-Aperture Radar}
\newacronym{CFAR}{CFAR}{Constant False Alarm Rate}
\newacronym{STTC}{STTC}{Space-Time Transmitting Code}
\newacronym{MIA}{MIA}{Majorized Iterative Algorithm}
\newacronym{MIACMC}{MIA-CMC}{Majorized Iterative Algorithm - Constant Modulus Constraint}
\newacronym{MIAPC}{MIA-PC}{Majorized Iterative Algorithm - PAR Constraint}
\newacronym{SVD}{SVD}{Single Value Decomposition}
\newacronym{UNIQUE}{UNIQUE}{UNImodular set of seQUEnce design}
\newacronym{WISE}{WISE}{Waveform design for beampattern shapIng and SpEctral masking}
\newacronym{SRR}{SRR}{Short-Range Radar}
\newacronym{MRR}{MRR}{Mid-Range Radar}
\newacronym{LRR}{LRR}{Long-Range Radar}
\newacronym{SILR}{SILR}{Spectral Integrated Level Ratio}
\newacronym{ED}{ED}{Eigenvalue Decomposition}
\newacronym{IBS}{IBS}{Iterative Beampattern with Spectral
design}
\newacronym{BIC}{BIC}{Beampattern Optimization With Spectral
Interference Control}

\newcommand{\ZEROV}{\mathbf{0}}

\newtheorem{theorem}{Theorem}[section]

\newtheorem{proof}[theorem]{proof}

\begin{document}
\title{MIMO Radar Transmit Beampattern Shaping for Spectrally Dense Environments}

\author{Ehsan~Raei,~\IEEEmembership{Student Member,~IEEE,}
        Saeid~Sedighi,~\IEEEmembership{Member,~IEEE,} Mohammad~Alaee-Kerahroodi,~\IEEEmembership{Member,~IEEE,}
        and~M.R.~Bhavani~Shankar,~\IEEEmembership{Senior Member,~IEEE}% <-this % stops a space
\thanks{This work was supported by \gls{FNR} through the BRIDGES project AWARDS under Grant CPPP17/IS/11827256/AWARDS and CORE project SPRINGER}}

\maketitle

\begin{abstract}
Designing unimodular waveforms with a desired beampattern, spectral occupancy and orthogonality level is of vital importance in the next generation Multiple-Input Multiple-Output (MIMO) radar systems. Motivated by this fact, in this paper, we propose a framework for shaping the beampattern in MIMO radar systems under the constraints simultaneously ensuring unimodularity, desired spectral occupancy and orthogonality of the designed waveform. In this manner, the proposed framework is the most comprehensive approach for MIMO radar waveform design focusing on beampattern shaping. The problem formulation leads to a non-convex quadratic fractional programming. We propose an effective iterative to solve the problem, where each iteration is composed of a Semi-Definite Programming (SDP) followed by eigenvalue decomposition. Some numerical simulations are provided to illustrate the superior performance of our proposed over the state-of-the-art.  
% In this paper, we design a set of waveforms for colocated Multiple-Input Multiple-Output (MIMO) radar systems to shape transmit beampattern. To this end, we consider spatial Integrated Sidelobe Level (ISLR) as the design metric and optimize it under continuous phase, spectral masking and similarity constraints. The problem formulation leads to a quadratic fractional programming, which is recast as a rank-one constrained optimization problem. We propose an effective iterative Semi-Definite Programming (SDP)-based method to solve the problem, where each iteration pushes the eigenvalues of the corresponding rank-one matrix except for its greatest eigenvalue to go to zero. Through numerical simulations, we illustrate the performance of our proposed method and compare it with state-of-the art. 
\end{abstract}


% \fi
\IEEEpeerreviewmaketitle

\section{Introduction}

% \IEEEPARstart{C}{ognitive} \gls{MIMO} radar systems are smart sensors which interact with the environment to enhance their performance \cite{guerci2010cognitive}. Spatial, spectrum and time are the most important resources, and resources management is one important aspects in cognitive \gls{MIMO} radar systems. This resources management necessitates an adaptive waveform design approach. Generally, beampattern shaping, spectral shaping and auto- and cross-correlation sidelobes minimization are the approaches which can be considered for managing the spatial, spectrum and time resources, respectively. In the following we enumerate the state of the arts in beampattern shaping, spectral shaping and auto- and cross-correlation sidelobes minimization via waveform design for \gls{MIMO} radar systems.

% \subsection{Beampattern Shaping} \label{subsec:Beampattern shaping} 
Transmit beampattern shaping by controlling the spatial distribution of the transmit power, can play an important role in improving the radar performance through enhanced power efficiency, better detection probability, target identification, improved interference mitigation, among others. The goal is to focus the transmit power on desired angles while minimizing it at undesired angles \cite{4217406}. Recently, the beampattern shaping via waveform design in \gls{MIMO} radar systems has been widely studied. From a waveform design perspective, there are two methods for beampattern shaping, indirect and direct methods \cite{7955071, 8713914}. In indirect (two-step) method, the waveform correlation matrix is firstly designed and the waveform matrix is subsequently obtained through one of the decomposition methods \cite{4276989, 4524058, 6649991, 7762192, 7126203, 4567663, 6747391, 7829401, 5765721} while in direct method, the waveform is designed in one step \cite{9440807, 9054519, 9104272, 7955071, 8387476, 8706630, 8713914, 8720029, 9466196}. On the other hand, there are several metrics (objective functions) to obtain the optimum beampattern such as, beampattern matching, spatial-\gls{ISLR}/\gls{PSLR} minimization, and \gls{SINR} maximization. 

\paragraph{\sl Beampattern Matching} In beampattern matching, the aim is to minimize the difference between the desired and designed beampattern. For instance, the following papers have worked on designing the waveform covariance matrix employing beampattern matching. The authors in \cite{4276989} devised a method to address the joint beampattern shaping and the cross-correlation minimization in spatial domain through \gls{SQP} technique. In \cite{4524058}, \gls{CA} is presented to shape the beampattern under low \gls{PAR} constraint. In \cite{6747391, 7829401}, the authors propose a covariance matrix design method based on \gls{DFT} coefficients and Toeplitz matrices. The \gls{DFT}-based technique provides a well-match transmit beampattern at low complexity. However, the drawback of the \gls{DFT}-based method is that, for small number of antennas, the performance of the \gls{DFT}-based method is slightly poorer. On the other hand, several papers focus on designing directly the transmit waveforms for beampattern shaping. For example, in \cite{7955071}, two optimization algorithms based on \gls{ADMM} are proposed under constant modulus constraint for the probing waveform. In \cite{8713914}, a method based in \gls{ADMM} is proposed to design a beampattern in wide-band systems. In \cite{8706630}, a method for beampattern matching is addressed based on gradient decent which they term it \gls{PDR}. In \cite{8720029}, the authors propose a method based on \gls{MM} for beampattern matching under \gls{PAR} constraint in two cases of wide- and narrow-band.

\paragraph{\sl Spatial-\gls{ISLR} and \gls{PSLR} minimization} In Spatial-\gls{ISLR} and \gls{PSLR} minimization approach, the aim is to minimize the ratio of {\it summation of beampattern response on undesired over desired angles}, and to minimize the ratio of {\it maximum beampattern response on undesired angles over minimum beampattern response on desired angles}, respectively. In \cite{7126203}, a method based on \gls{SDR} under constant energy and $3$ dB main beam-width constraint is proposed to minimize the spatial-\gls{ISLR}. In \cite{7435338}, the robust designs of waveform covariance matrix through optimizing the worst case transmit beampattern are considered to minimize the spatial-\gls{ISLR} and -\gls{PSLR} of beampatterns, respectively. Unlike two aforementioned methods, \cite{9440807, 9466196, 8387476} propose a direct waveform design solution. The authors in \cite{9440807} propose the efficient \gls{UNIQUE} algorithm based on \gls{CD} method to minimize spatial- and range-\gls{ISLR} under four different constraints, namely, limited energy, \gls{PAR}, continuous and discrete phase constraints. The method proposed in \cite{9466196} is similar to \gls{UNIQUE} without considering range-\gls{ISLR} metric and \gls{PAR} and limited energy constraints. A method based on \gls{ADMM} is proposed in \cite{8387476} to minimize the spatial-\gls{PSLR} under constant modulus constraint.

\paragraph{\sl \gls{SINR} maximization} In \gls{SINR} optimization approaches, the problem does not deal with the beampattern directly. However, the beampattern is implicitly shaped as a result of transmit waveform optimization. For example \cite{6649991, 7762192} address the problem of waveform design in the presence of signal dependence clutter. In these works, an iterative approach is presented to jointly optimize the transmit waveform and receive filter to maximize the output \gls{SINR}. The authors in \cite{8239836} propose \gls{MIA} based on \gls{MM} method for joint waveform and filter design under similarity, constant modulus (\gls{MIACMC}) and \gls{PAR} (\gls{MIAPC}) constraints. While \gls{STTC} \cite{8401959} is proposed based on \gls{CD} to solve the problem under similarity, uncertain steering matrices, continuous or discrete phase constraints. In \cite{8401959}, a Dinkelbach based method and exhaustive search is proposed for continuous and discrete phase constraints respectively.

In order to form the virtual array and enhancing the angular resolution, the received signal in \gls{MIMO} radar system should be separable (orthogonal) in receiver while a set of arbitrary waveforms are adopted in the transmit side. In order to obtain the orthogonality, the waveform should have small cross-correlation \cite{LImimo}. Also, small auto-correlation sidelobes is a requirement, to avoid masking weak targets within the range sidelobes of a strong target, and to mitigate the harmful effects of distributed clutter returns close to the target of interest. Recently, many optimization techniques, e.g., Multi-\gls{CAN} \cite{5072243,he2012waveform}, Iterative Direct Search \cite{cui2017constant}, \gls{ISL} New \cite{7760646,8239862}, \gls{MM}-Corr \cite{7420715}, \gls{BiST} \cite{8706639}, \gls{UNIQUE} \cite{9440807} and \gls{WeBEST} \cite{raei2021design} are proposed to design orthogonal sets of sequences, minimizing the \gls{ISL}/\gls{PSL} metrics. However, beampattern shaping in \gls{MIMO} radar systems yield a correlated waveform which, is in contradiction with orthogonality \cite{9440807, 4516997}. In this context there are few papers which addressed these two aspects in \gls{MIMO} radar systems. For instance \cite{4516997} proposes beampattern matching problem under particular constraints on the waveform cross-correlation matrix. In \cite{7511868}, the authors minimizes the difference between desired and undesired beampattern responses for one sub-pulse. Then the quasi-orthogonality of other sub-pulses are obtained by random permutation. In \cite{8645373}, the authors combine a beampattern matching by orthogonality requirement as a penalty in the objective function and use the \gls{PDR} approach for the solution. In \cite{8835646}, the authors propose a method based on \gls{ADMM} to design a beampattern with good cross-correlation. In \cite{9440807}, \gls{UNIQUE} is proposed as a unified framework to minimize the spatial- and range-\gls{ISLR} using weighted sum technique under limited energy, \gls{PAR}, continuous and discrete phase constraints. By choosing an appropriate value for the regularization parameter, \gls{UNIQUE} is able to make trade-off between having a good orthogonality and beampattern shaping. 

On the other hand, spectral shaping is an important aspect of resource management in cognitive\footnote{Cognitive \gls{MIMO} radar systems are smart sensors which interact with the environment to adapt the properties of the waveform with the environment to enhance their performance \cite{guerci2010cognitive}.} \gls{MIMO} radar systems. Uing this approach, the cognitive radar system is able to utilize effectively the available bandwidth. One attractive application of spectral shaping is coexistence of communications and cognitive \gls{MIMO} radar systems, which the whole bandwidth is shared between these two systems based on the priorities \cite{alaeekerahroodi2021coexistence}. There are several methods for spectral shaping. For instance, in \cite{5604089,6784117,7174964,7529179,9122033,9337317} spectral matching approach is proposed to shape the spectral of the transmit waveform. In \cite{8579200,9052442}, the authors consider a waveform design approach to maximize \gls{SINR}, while the spectral behaviour is considered as a constraint. In \cite{8358735,8770133}, the ratio of the maximum stop-band level to the minimum pass-band level is considered as the objective function to shape the spectrum. \gls{SILR} minimization approach is consider under continuous and discrete phase constraints in \cite{alaeekerahroodi2021coexistence}. The design of constant modulus waveform for beampattern matching under spectral constraint are addressed in \cite{9122033, 8429253}. To tackle the non-convex optimization problem the authors in \cite{9122033} and \cite{8429253} propose \gls{IBS} and \gls{BIC} methods respectively.

\subsection{Contribution}
In this paper we consider the spatial-\gls{ISLR} as design metric similar to \cite{9466196}. In \cite{9466196}, the authors proposed \gls{CD}-based method to enhance the performance of the radar in spatial domain. However, in this paper, we deal with the design of waveform considering the features in three domains: \gls{ISLR} in the spatial and range domain and masking in the spectral domain.
% we propose a new framework for shaping the beampattern in \gls{MIMO} radar systems using spatial-\gls{ISLR} as the design metric under the constraints of simultaneously ensuring unimodularity, desired spectral occupancy and orthogonality of the designed waveform.
Particularly, we propose a waveform design framework to shape the beampattern with practical constraints, namely, spectral masking, $3$ dB beam-width, constant modulus and similarity constraints. 
Spectral masking constraint plays an important role in cognitive \gls{MIMO} radar systems in several scenarios, such as spectral sharing in coexistence of \gls{MIMO} radar and \gls{MIMO} communication.
The $3$ dB beam-width constraint ensures that the beampattern has a good response at the mainlobe. This constraint can be used in the emerging 4D-imaging automotive \gls{MIMO} radar systems, wherein the \gls{SRR}, \gls{MRR} and \gls{LRR} configurations are merged, to provide unique and high angular resolution in the entire radar detection range \cite{9440807, 9466196}.
Constant modulus waveforms are attractive for radar system designers due to efficient utilization of the limited transmitter power. Besides, since constant modulus is a kind of only phase-modulated sequence, implementing of constant modulus waveform is simple.
As to the orthogonality, we incorporate the beampattern shaping optimization problem with similarity constraints to make a trade-off between having a good beampattern response and orthogonality \cite{9440807}. This constraint imposes that the optimize waveform inherit some properties of a reference waveform. In fact, we consider the designed waveform to be similar to a specific waveform which have a good orthogonality to form the virtual array in receivers and enhance its angular resolution.

It is desirable to include all these properties to improve radar performance in emerging applications and in the emerging scenario of crowded environments with interference from other radars or communication systems. In this context, the contributions of the work are listed below.
% In this manner, the proposed framework is the most comprehensive approach for MIMO radar waveform design focusing on beampattern shaping.
% In this paper, we consider the transmit beampattern shaping problem for colocated \gls{MIMO} radar systems via waveform design. 
% We consider an optimization approach to minimize the spatial-\gls{ISLR} as design metric. The spatial-\gls{ISLR} metric obtains better nulls in compare with beampattern matching approaches, while the beampattern matching offers better mainlobe response. We consider practical constraints, which play an important role in enhancing the performance of \gls{MIMO} radar systems.
% In this paper we propose a waveform design approach for beampattern shaping in \gls{MIMO} radar systems. We consider an optimization approach to minimize the spatial-\gls{ISLR} as design metric. The spatial-\gls{ISLR} metric obtains better nulls in compare with beampattern matching approaches, while the beampattern matching offers better mainlobe response. 
% The main contributions of this paper are summarized below.
\begin{itemize}
\item {\sl Incorporation of metrics from multiple domains:} Radar tasks are influenced by parameters in the spatial, temporal and spectral domain. Hence it is pertinent to consider quality metrics in all these domains to improve performance. Thus, while it is highly interesting to consider all the metrics in the waveform design, the existing works consider only a selection of these performance metrics. A problem set-up involving these key performance indicators in different domains is lacking in literature. In this context the proposed framework incorporating all the metrics is the most comprehensive approach for MIMO radar waveform design focusing on beampattern shaping; it subsumes existing works as special cases. The gains obtained by incorporating these metrics over the existing works bears testimony to their impact.

\item {\sl Novel optimization framework:} 
The incorporation of all the aforementioned quality metrics adds further complexity to the waveform optimization and these cannot be handled by the existing frameworks. In this context, the paper also offers a novel optimization framework to solve the non-convex multi-variable and NP-hard problem. In an attempt to solve this problem, we propose an indirect method based on \gls{SDP}. We first recast the waveform-design problem as a rank-one constrained optimization problem. Then, unlike the conventional methods which drop the rank one constraint, we propose a new iterative algorithm for efficiently solving the resulting rank-one constrained optimization problem. Each iteration of the proposed iterative algorithm is composed of an \gls{SDP} followed by an \gls{ED}. In every iteration, we force the second largest eigen value towards zero to obtain the rank one solution. We prove that the proposed iterative algorithm converges to a local minima of the rank-one constrained optimization problem. Further, we compute the computational complexity of the proposed iterative algorithm. In addition the proposed framework can be extended to apply other convex constraints.

% In this paper, we first recast the waveform-design problem as a rank-one constrained optimization problem. Then, we propose a new iterative algorithm for efficiently solving the resulting rank-one constrained optimization problem. Each iteration of the proposed iterative algorithm is composed of an Semi-Definite Programming (SDP) followed by an Eigenvalue Decomposition (ED). We prove that the proposed iterative algorithm converges to a local minima of the rank-one constrained optimization problem. Further, we compute the computational complexity of the proposed iterative algorithm.
\end{itemize}

\subsection{Organization and Notation}
The rest of this research is organized as follows. In Section \ref{sec:System Model and Problem Formulation}, the system model and the design problem for minimizing the spatial-\gls{ISLR} under constant modulus, $3$ dB beam-width, similarity and spectral masking, constraints is formulated. We develop the iterative \gls{WISE} framework based on \gls{SDP} to obtain a rank-one solution in Section \ref{sec:Proposed method}. Finally we provide numerical experiments to verify the effectiveness of proposed algorithm in Section \ref{sec:Numerical Results}.
\paragraph*{Notations} This paper uses lower-case and upper-case boldface for vectors ($\mathbf{a}$) and matrices ($\mathbf{A}$) respectively. The conjugate, transpose and the conjugate transpose operators are denoted by the $(.)^*$, $(.)^T$ and $(.)^{\dagger}$ symbols respectively. Besides the Frobenius norm, $\ell_2$ and $\ell_p$ norm, absolute value and round operator are denoted by $\norm{.}_F$, $\norm{.}_2$, $\norm{.}_p$, $|.|$ and $\lfloor . \rceil$ respectively. For any matrix $\mathbf{A}$, $\text{Tr}(\mathbf{A})$, $\text{Diag}(\mathbf{A})$ and $\text{Rank}(\mathbf{A})$ stand for the trace, diagonal vector and the rank of $\mathbf{A}$, respectively. 
% For any complex number $a$, $\Re(a)$ and $\Im(a)$ denotes the real and imaginary part, respectively. 
The letter $j$ represents the imaginary unit (i.e., $j=\sqrt{-1}$), while the letter $(i)$ is use as step of a procedure. Finally $\mathbf{1}$ and $\ZEROV$ denote a matrix/vector with proper size which all the elements are equal to one and zero respectively. 
% Finally $\odot$ and $\mathbf{1}_M$ denote the Hadamard product and a $M\times1$ vector which al the elements are equal to one, respectively. 

\section{System Model}\label{sec:System Model and Problem Formulation}
We consider a colocated narrow-band \gls{MIMO} radar system, with $M$ transmit antennas, each transmitting a sequence of length $N$ in the fast-time domain. Let the matrix $\mathbf{S} \in \mathbb{C}^{M \times N}$ denote the transmitted set of sequences in baseband, i.e,
\begin{equation*}\label{eq:S}
	\mathbf{S} \triangleq 
	\begin{bmatrix}
	s_{1,1} & s_{1,2} & \dots  & s_{1,N} \\
	s_{2,1} & s_{2,2} & \dots  & s_{2,N} \\
	\vdots  & \vdots  & \vdots & \vdots  \\
	s_{M,1} & s_{M,2} & \dots  & s_{M,N}
	\end{bmatrix}.
\end{equation*}

Let us further denote that $\mathbf{S} \triangleq [\bar{\mathbf{s}}_1, \dots, \bar{\mathbf{s}}_N] \triangleq [\tilde{\mathbf{s}}_1^T, \dots, \tilde{\mathbf{s}}_M^T]^T$, where the vector $\bar{\mathbf{s}}_n \triangleq [s_{1,n}, s_{2,n}, \ldots, s_{M,n}]^T \in \mathbb{C}^{M}$ ($n=\{1,\dots,N\}$) indicates the $n^{th}$ time-sample across the $M$ transmitters (the $n^{th}$ column of matrix $\mathbf{S}$) while the $\tilde{\mathbf{s}}_m \triangleq [s_{m,1}, s_{m,2}, \dots, s_{m,N}]^T \in \mathbb{C}^N$ ($m=\{1,\dots,M\}$) indicates the $N$ samples of $m^{th}$ transmitter (the $m^{th}$ row of matrix $\mathbf{S}$). In this paper, 
% we deal with the spatial-\gls{ISLR}, spectrum masking and similarity constraints. To this end, in the following, we introduce the system model in these two domains.
we deal with the design of $\mathbf{S}$ considering features in three domains: \gls{ISLR} in the spatial and range domain and masking in the spectral domain. To this end, we now introduce the system model to describe in spatial and spectral domains. Subsequently, we also introduce similarity constraints to impose the range-\gls{ISLR} features.

% Let, $\mathbf{s}_n \triangleq [s_{1,n}, s_{2,n}, \ldots, s_{M,n}]^T \in \mathbb{C}^{M}$ ($n=\{1,\dots,N\}$) be the $n^{th}$ time-sample across the $M$ transmitters (the $n^{th}$ column of matrix $\mathbf{S}$ ($\mathbf{S} \triangleq [\mathbf{s}_1, \dots, \mathbf{s}_N]$)). 

\subsection{System Model in Spatial Domain}
Let assume a colocated \gls{MIMO} radar system with an \gls{ULA} structure for the transmit array characterized by the following steering vector\cite{LImimo},
\begin{equation}\label{eq:a(theta)}
	\mathbf{a}(\theta)=[1, e^{j\frac{2\pi d}{\lambda} \sin(\theta)}, \ldots, e^{j\frac{2\pi d(M-1)}{\lambda} \sin(\theta)}]^T \in \mathbb{C}^M,
\end{equation}
where $d$ is the distance between the transmitter antennas and $\lambda$ is the signal wavelength. The beampattern in the direction of $\theta$ can be written as \cite{LImimo,7435338,4276989},
\begin{equation*}\label{eq:beampattern}
	P(\mathbf{S}, \theta) = \frac{1}{N}\sum_{n=1}^{N} \left|\mathbf{a}^{{\dagger}}(\theta) \bar{\mathbf{s}}_n\right|^2 = \frac{1}{N}\sum_{n=1}^{N}\bar{\mathbf{s}}_n^{\dagger}\mathbf{A}(\theta)\bar{\mathbf{s}}_n
\end{equation*}
where, $\mathbf{A}(\theta) = \mathbf{a}(\theta)\mathbf{a}^{\dagger}(\theta)$. 
Let $\Theta_d$ and $\Theta_u$ be the sets of desired and undesired angles in the spatial domain, respectively. These two sets satisfy, $\Theta_d \cap \Theta_u = \varnothing$ and $\Theta_d \cup \Theta_u \subset [-90^o, 90^o]$. In this regard the spatial-\gls{ISLR} is given by the following expression \cite{9440807},
\begin{equation}\label{eq:Spatial_ISLR}
	f(\mathbf{S}) \triangleq \frac{\sum_{\theta \in \Theta_u} P(\mathbf{S},\theta)}{\sum_{\theta \in \Theta_d} P(\mathbf{S},\theta)} = \frac{\sum_{n=1}^{N}\bar{\mathbf{s}}_n^{\dagger}\mathbf{A}_u\bar{\mathbf{s}}_n}{\sum_{n=1}^{N}\bar{\mathbf{s}}_n^{\dagger}\mathbf{A}_d\bar{\mathbf{s}}_n},
\end{equation} 
where $\mathbf{A}_u \triangleq \frac{1}{N}\sum_{\theta \in \Theta_u}\mathbf{A}(\theta)$ and $\mathbf{A}_d \triangleq \frac{1}{N}\sum_{\theta \in \Theta_d}\mathbf{A}(\theta)$. 

\subsection{System Model in Spectrum Domain}
Let $\mathbf{F} \triangleq [\mathbf{f}_0,\dots,\mathbf{f}_{N-1}] \in \mathbb{C}^{N \times N} $ be the \gls{DFT} matrix, where, $\mathbf{f}_k \triangleq [1, e^{-j\frac{2 \pi k}{N}}, \dots, e^{-j\frac{2 \pi k(N-1)}{N}}]^T \in \mathbb{C}^N, \ k = \{0, \dots, N-1\}$. Let $\mathcal{U} = \cup_{k=1}^{K_u}(u_{k,1}, u_{k,2})$ be the $K_u$ number of normalized frequency stop-bands, where $0 \leq u_{k,1} < u_{k,2} \leq 1$ and $\cap_{k=1}^{K_u}(u_{k,1}, u_{k,2}) = \varnothing$. Thus, the undesired discrete frequency bands are given by $\mathcal{V} = \cup_{k=1}^{K_u}(\lfloor N u_{k,1}\rceil, \lfloor N u_{k,2}\rceil)$.
% Let $\mathcal{U} \subset \{0,\dots,N-1\}$ be the stop discrete frequency bands for \gls{MIMO} radar. 
In this regards the 
% power spectrum at undesired frequency bins 
absolute value of the spectrum an undesired frequency bins 
can be expressed as $|\mathbf{G}\tilde{\mathbf{s}}_m|$, where, $\mathbf{G} \in \mathbb{C}^{K \times N}$ contains the rows of $\mathbf{F}$ corresponding to the frequencies in $\mathcal{V}$, and $K$ is the number of undesired frequency bins \cite{alaeekerahroodi2021coexistence}.

% as $\left|\mathbf{f}_k^{\dagger} \tilde{\mathbf{s}}_m\right| | k \in \mathcal{U} $ \cite{alaeekerahroodi2021coexistence}.

\subsection{Problem Formulation}\label{subsec:Problem Formulation}
We aim to design a set of constant modulus sequences for \gls{MIMO} radar such that the transmit beampattern is steered towards desired directions and has nulls at undesired directions simultaneously, with spectrum compatibility and similarity constraints. To this end, we can formulate the following optimization problem,
% \par\noindent\small
% \begin{subequations}\label{eq:P0}
%   \begin{empheq}[left=\empheqlbrace]{align}
%     \label{eq:P0_a} \min_{\mathbf{S}} & \quad f(\mathbf{S}) = \frac{\sum_{n=1}^{N}\bar{\mathbf{s}}_n^{\dagger}\mathbf{A}_u\bar{\mathbf{s}}_n}{\sum_{n=1}^{N}\bar{\mathbf{s}}_n^{\dagger}\mathbf{A}_d\bar{\mathbf{s}}_n}\\ 
%     \label{eq:P0_b} s.t. & \quad 0.5 \leq \frac{\sum_{n=1}^{N}\bar{\mathbf{s}}_n^{\dagger}\mathbf{A}(\theta)\bar{\mathbf{s}}_n} {\sum_{n=1}^{N}\bar{\mathbf{s}}_n^{\dagger}\mathbf{A}(\theta_0)\bar{\mathbf{s}}_n} \leq 1, ~ \forall \theta \in \Theta_d,\\
%     \label{eq:P0_c} & \quad \max \left\{\left|\mathbf{f}_k^{\dagger} \tilde{\mathbf{s}}_m\right|\right\} \leq \gamma ~ | k \in \mathcal{U}, ~ m\in\{1,\dots,M\} \\
%     \label{eq:P0_d} & \quad |s_{m,n}| = 1, \\ 
%     \label{eq:P0_e} & \quad \frac{1}{\sqrt{MN}}\norm{\mathbf{S} - \mathbf{S}_0}_F \leq \delta
%   \end{empheq}
% \end{subequations}
% \normalsize

% \par\noindent\small
\begin{subequations}\label{eq:P0}
  \begin{empheq}[left=\empheqlbrace]{align}
    \label{eq:P0_a} \min_{\mathbf{S}} & \quad f(\mathbf{S}) = \frac{\sum_{n=1}^{N}\bar{\mathbf{s}}_n^{\dagger}\mathbf{A}_u\bar{\mathbf{s}}_n}{\sum_{n=1}^{N}\bar{\mathbf{s}}_n^{\dagger}\mathbf{A}_d\bar{\mathbf{s}}_n}\\ 
    \label{eq:P0_b} s.t. & \quad 0.5 \leq \frac{\sum_{n=1}^{N}\bar{\mathbf{s}}_n^{\dagger}\mathbf{A}(\theta_d)\bar{\mathbf{s}}_n} {\sum_{n=1}^{N}\bar{\mathbf{s}}_n^{\dagger}\mathbf{A}(\theta_0)\bar{\mathbf{s}}_n} \leq 1,\\
    \label{eq:P0_c} & \quad |s_{m,n}| = 1, \\
    \label{eq:P0_d} & \quad \max \left\{\left|\mathbf{G}\tilde{\mathbf{s}}_m\right|\right\} \leq \gamma, ~ m\in\{1,\dots,M\}, \\ 
    \label{eq:P0_e} & \quad \frac{1}{\sqrt{MN}}\norm{\mathbf{S} - \mathbf{S}_0}_F \leq \delta,
  \end{empheq}
\end{subequations}
% \normalsize
where \eqref{eq:P0_b} indicates the $3$ dB beam-width constraint to guarantee the beampattern response at all desired angles is at least half the maximum power. In \eqref{eq:P0_b}, $\theta_d \in \{\theta| \forall \theta \in \Theta_d\}$ and $\theta_0$, denotes the the angle with maximum power, which is usually chosen to be at the center point of $\Theta_d$. The constraint \eqref{eq:P0_c} indicates the constant modulus property; this is attractive for radar system designers since its allows for the efficient utilization of the limited transmitter power. The constraint \eqref{eq:P0_d} indicates the spectrum masking and guarantees the power of spectrum in undesired frequencies not to be greater than $\gamma$. Finally, the constraint \eqref{eq:P0_e} has been imposed on the waveform to control properties of the optimized code (such as orthogonality) similar to the reference waveform $\mathbf{S}_0$, for instance this helps controlling \gls{ISLR} in range domain. If we consider $\mathbf{S}$ and $\mathbf{S}_0$ to be a constant modulus waveform, the maximum admissible value of the similarity constraint parameter would be $\delta = \sqrt{2}$ ($0 \leq \delta \leq \sqrt{2}$).

In \eqref{eq:P0}, the objective function \eqref{eq:P0_a} is a fractional quadratic function, \eqref{eq:P0_b} and \eqref{eq:P0_d} are non-convex inequality constraints. The \eqref{eq:P0_c} is a non-affine equality constraint while, the inequality constraint \eqref{eq:P0_e} yields a convex set. Therefore, we encounter a non-convex, multi-variable and NP-hard optimization problem \cite{9440807}. In the following, we propose an iterative method based on \gls{SDP} to solve the problem efficiently.

\section{Proposed Method}\label{sec:Proposed method}
The maximum of $P(\mathbf{S},\theta)$ is clearly $M^2$, and occurs when $\bar{\mathbf{s}}_n = \mathbf{a}(\theta)$ $n = \{1, \ldots, N\}$. Therefore, the denominator of \eqref{eq:P0_a} satisfies, $\sum_{n=1}^{N}\bar{\mathbf{s}}_n^{\dagger}\mathbf{A}_d\bar{\mathbf{s}}_n \leq K_d M^2$, where $K_d$ is the number of desired angles. Thus, the problem \eqref{eq:P0} can be equivalently written as \cite{8314727},

% \begin{subequations}\label{eq:P1}
%   \begin{empheq}[left=\empheqlbrace]{align}
%     \label{eq:P1_a} \min_{\mathbf{S}} & \quad  \sum_{n=1}^{N}\bar{\mathbf{s}}_n^{\dagger}\mathbf{A}_u\bar{\mathbf{s}}_n \\ 
%     \label{eq:P1_b} s.t. & \quad \sum_{n=1}^{N}\bar{\mathbf{s}}_n^{\dagger}\mathbf{A}_d\bar{\mathbf{s}}_n \leq K_d M^2, \\
%     \label{eq:P1_c} & \quad  \sum_{n=1}^{N}\bar{\mathbf{s}}_n^{\dagger}\mathbf{A}(\theta)\bar{\mathbf{s}}_n  \leq \sum_{n=1}^{N}\bar{\mathbf{s}}_n^{\dagger}\mathbf{A}(\theta_0)\bar{\mathbf{s}}_n, ~ \forall \theta \in \Theta_d,\\
%     \label{eq:P1_d} & \quad \sum_{n=1}^{N}\bar{\mathbf{s}}_n^{\dagger}\mathbf{A}(\theta_0)\bar{\mathbf{s}}_n  \leq 2\sum_{n=1}^{N}\bar{\mathbf{s}}_n^{\dagger}\mathbf{A}(\theta)\bar{\mathbf{s}}_n, ~ \forall \theta \in \Theta_d,\\
%     \label{eq:P1_e} & \quad \norm{\mathbf{f}_k^{\dagger} \tilde{\mathbf{s}}_m}_{p\to\infty} \leq \gamma ~ | k \in \mathcal{U}, ~ m\in\{1,\dots,M\} \\
%     \label{eq:P1_f}  & \quad |s_{m,n}| = 1, \\
%     \label{eq:P1_g} & \quad \frac{1}{\sqrt{MN}}\norm{\mathbf{S} - \mathbf{S}_0}_F \leq \delta
%   \end{empheq}
% \end{subequations}

\begin{subequations}\label{eq:P1}
  \begin{empheq}[left=\empheqlbrace]{align}
    \label{eq:P1_a} \min_{\mathbf{S}} & \quad  \sum_{n=1}^{N}\bar{\mathbf{s}}_n^{\dagger}\mathbf{A}_u\bar{\mathbf{s}}_n \\ 
    \label{eq:P1_b} s.t. & \quad \sum_{n=1}^{N}\bar{\mathbf{s}}_n^{\dagger}\mathbf{A}_d\bar{\mathbf{s}}_n \leq K_d M^2, \\
    \label{eq:P1_c} & \quad  \sum_{n=1}^{N}\bar{\mathbf{s}}_n^{\dagger}\mathbf{A}(\theta_d)\bar{\mathbf{s}}_n  \leq \sum_{n=1}^{N}\bar{\mathbf{s}}_n^{\dagger}\mathbf{A}(\theta_0)\bar{\mathbf{s}}_n,\\
    \label{eq:P1_d} & \quad \sum_{n=1}^{N}\bar{\mathbf{s}}_n^{\dagger}\mathbf{A}(\theta_0)\bar{\mathbf{s}}_n  \leq 2\sum_{n=1}^{N}\bar{\mathbf{s}}_n^{\dagger}\mathbf{A}(\theta_d)\bar{\mathbf{s}}_n,\\
    \label{eq:P1_e} & \quad |s_{m,n}| = 1, \\
    \label{eq:P1_f}  & \quad \norm{\mathbf{G}\tilde{\mathbf{s}}_m}_{p\to\infty} \leq \gamma, ~ m\in\{1,\dots,M\}, \\
    \label{eq:P1_g} & \quad \frac{1}{\sqrt{MN}}\norm{\mathbf{S} - \mathbf{S}_0}_F \leq \delta
  \end{empheq}
\end{subequations}

In \eqref{eq:P1}, constraints \eqref{eq:P1_c} and \eqref{eq:P1_d} are obtained by expanding \eqref{eq:P0_b} constraint. Besides, we replace the non-convex constraint $\max\{|\mathbf{G}\tilde{\mathbf{s}}_m|\}$ \eqref{eq:P0_d} with $\norm{\mathbf{G}\tilde{\mathbf{s}}_m}_{p\to\infty}$ \eqref{eq:P1_f}, 
which is a convex constraint for each finite $p$. 

Problem \eqref{eq:P1} is 
% a separable problem in $\bar{\mathbf{s}}_1, \bar{\mathbf{s}}_2, \cdots, \bar{\mathbf{s}}_N$ but it is 
still non-convex with respect to $\mathbf{S}$ due to \eqref{eq:P1_e}. To cope with this problem, defining $\mathbf{X}_n \triangleq \bar{\mathbf{s}}_n\bar{\mathbf{s}}_n^{\dagger}$, we recast \eqref{eq:P1} as follows:
\begin{subequations}\label{eq:P2}
  \begin{empheq}[left=\empheqlbrace]{align}
    \label{eq:P2_a} \min_{\mathbf{S}, \mathbf{X}_n} & \quad  \sum_{n=1}^{N}\text{Tr}(\mathbf{A}_u\mathbf{X}_n) \\ 
    \label{eq:P2_b} s.t. & \quad \sum_{n=1}^{N}\text{Tr}(\mathbf{A}_d\mathbf{X}_n) \leq K_d M^2, \\
    \label{eq:P2_c} & \quad   \sum_{n=1}^{N}\text{Tr}(\mathbf{A}(\theta_d)\mathbf{X}_n)  \leq \sum_{n=1}^{N}\text{Tr}(\mathbf{A}(\theta_0)\mathbf{X}_n),\\
    \label{eq:P2_d} & \quad \sum_{n=1}^{N}\text{Tr}(\mathbf{A}(\theta_0)\mathbf{X}_n)   \leq 2\sum_{n=1}^{N}\text{Tr}(\mathbf{A}(\theta_d)\mathbf{X}_n),\\
    \label{eq:P2_e}  & \quad \text{Diag}(\mathbf{X}_n) = \mathbf{1}_M, \\
    \label{eq:P2_f} & \quad \eqref{eq:P1_f}, \eqref{eq:P1_g}, \\
    \label{eq:P2_g} & \quad \mathbf{X}_n \succcurlyeq \ZEROV,\\
    \label{eq:P2_h} & \quad \mathbf{X}_n = \bar{\mathbf{s}}_n\bar{\mathbf{s}}_n^{\dagger},
    % \label{eq:P2_i} & \quad \text{Rank}(\mathbf{X}_n) = 1.
  \end{empheq}
\end{subequations}
It is readily observed that, in \eqref{eq:P2}, the objective function and all the constraints but \eqref{eq:P2_h} are convex in $\mathbf{X}_n$ and $\mathbf{S}$.
% A possible solution to tackle with \eqref{eq:P2} is relaxing the rank constraint by dropping the $\text{Rank}(\mathbf{X}_n) = 1$ constraint, which is known as an \gls{SDR} of \eqref{eq:P2} \cite{5447068}. 
% Solving \eqref{eq:P2} by \gls{SDR}, leads to obtain the optimum $\mathbf{X}_n$ and there exist several methods to approximate $\bar{\mathbf{s}}_n$, e.g., eigen-decomposition and Gaussian randomization. However, \gls{SDR} method does not guarantee to obtain a unique solution. 
In the following, we first present an equivalent reformulation for \eqref{eq:P2}, which paves the way for iteratively solving this non-convex optimization problem .
% to obtain a rank one solution.

\begin{theorem} \label{theorehm:t1}
Defining $
\mathbf{Q}_n \triangleq \begin{bmatrix}
1 & \bar{\mathbf{s}}_n^{\dagger} \\
\bar{\mathbf{s}}_n & \mathbf{X}_n
 \end{bmatrix} \in \mathbb{C}^{(M+1) \times (M+1)}
 $ and considering slack variables $\mathbf{V}_n \in \mathbb{C}^{(M+1)\times M}$ and $b_n \in \mathbb{R}$. The optimization problem \eqref{eq:P2} is equivalent to,
%
% \par\noindent\small
% \begin{subequations}\label{eq:P3}
%   \begin{empheq}[left=\empheqlbrace]{align}
%     \label{eq:P3_a} \min_{\mathbf{S}, \mathbf{X}_n, b_n} & \quad  \sum_{n=1}^{N}\text{Tr}(\mathbf{A}_u\mathbf{X}_n) + \eta \sum_{n=1}^{N}b_n\\ 
%     \label{eq:P3_b} s.t. & \quad \sum_{n=1}^{N}\text{Tr}(\mathbf{A}_d\mathbf{X}_n) \leq K_d M^2, \\
%     \label{eq:P3_c}  & \quad \text{Diag}(\mathbf{X}_n) = \mathbf{1}_M, \\ 
%     \label{eq:P3_d} & \quad   \sum_{n=1}^{N}\text{Tr}(\mathbf{A}(\theta)\mathbf{X}_n)  \leq \sum_{n=1}^{N}\text{Tr}(\mathbf{A}(\theta_0)\mathbf{X}_n)  ~ |  \forall \theta \in \Theta_d,\\
%     \label{eq:P3_e} & \quad \sum_{n=1}^{N}\text{Tr}(\mathbf{A}(\theta_0)\mathbf{X}_n)   \leq 2\sum_{n=1}^{N}\text{Tr}(\mathbf{A}(\theta)\mathbf{X}_n)  ~ | \forall \theta \in \Theta_d,\\
%     \label{eq:P3_f} & \quad \frac{1}{\sqrt{MN}}\norm{\mathbf{S} - \mathbf{S}_0}_F \leq \delta\\
%     \label{eq:P3_g} & \quad \mathbf{X}_n \succcurlyeq 0,\\
%     \label{eq:P3_h} & \quad \mathbf{Q}_n \succcurlyeq 0,\\
%     \label{eq:P3_i} & \quad b_n\mathbf{I}_n - \mathbf{V}_n^{\dagger}\mathbf{Q}_n\mathbf{V}_n \succcurlyeq 0,\\
%     \label{eq:P3_j} & \quad b_n \geq 0,
%   \end{empheq}
% \end{subequations}
% \normalsize

%
\begin{subequations}\label{eq:P3}
  \begin{empheq}[left=\empheqlbrace]{align}
    \label{eq:P3_a} \min_{\mathbf{S}, \mathbf{X}_n, b_n} & \quad  \sum_{n=1}^{N}\text{Tr}(\mathbf{A}_u\mathbf{X}_n) + \eta \sum_{n=1}^{N}b_n\\ 
    \label{eq:P3_b} s.t. & \quad \eqref{eq:P2_b}, \eqref{eq:P2_c}, \eqref{eq:P2_d}, \eqref{eq:P2_e}, \eqref{eq:P2_f}, \eqref{eq:P2_g}\\
    \label{eq:P3_h} & \quad \mathbf{Q}_n \succcurlyeq 0,\\
    \label{eq:P3_i} & \quad b_n\mathbf{I}_M - \mathbf{V}_n^{\dagger}\mathbf{Q}_n\mathbf{V}_n \succcurlyeq \ZEROV,\\
    \label{eq:P3_j} & \quad b_n \geq 0,
  \end{empheq}
\end{subequations}
where $\eta$ is a regularization parameter.
\end{theorem}

\begin{proof}
See Appendix \ref{app:1}.
\end{proof}

The problem \eqref{eq:P3} can be solved iteratively by alternating between the parameters. Let, $\mathbf{V}_n^{(i)}$, $\mathbf{Q}_n^{(i)}$, $\mathbf{S}^{(i)}$, $\mathbf{X}_n^{(i)}$ and $b_n^{(i)}$ be the values of $\mathbf{V}_n$, $\mathbf{Q}_n$, $\mathbf{S}$, $\mathbf{X}_n$ and $b_n$ at $i^{th}$ iteration, respectively. Given $\mathbf{V}^{(i-1)}$ and $b_n^{(i-1)}$, the optimization problem with respect to $\mathbf{S}^{(i)}$, $\mathbf{X}_n^{(i)}$ and $b_n^{(i)}$ at the $i^{th}$ iteration becomes,

\par\noindent\small
\begin{subequations}\label{eq:P5}
  \begin{empheq}[left=\empheqlbrace]{align}
    \label{eq:P5_a} \min_{\mathbf{S}^{(i)}, \mathbf{X}_n^{(i)}, b_n^{(i)}} & \quad  \sum_{n=1}^{N}\text{Tr}(\mathbf{A}_u\mathbf{X}_n^{(i)}) + \eta \sum_{n=1}^{N}b_n^{(i)}\\ 
    \label{eq:P5_b} s.t. & \quad \sum_{n=1}^{N}\text{Tr}(\mathbf{A}_d\mathbf{X}_n^{(i)}) \leq K_d M^2, \\
    \label{eq:P5_c} & \quad   \sum_{n=1}^{N}\text{Tr}(\mathbf{A}(\theta)\mathbf{X}_n^{(i)}) \leq \sum_{n=1}^{N}\text{Tr}(\mathbf{A}(\theta_0)\mathbf{X}_n^{(i)}),\\
    \label{eq:P5_d} & \quad \sum_{n=1}^{N}\text{Tr}(\mathbf{A}(\theta_0)\mathbf{X}_n^{(i)}) \leq 2\sum_{n=1}^{N}\text{Tr}(\mathbf{A}(\theta)\mathbf{X}_n^{(i)}),\\
    \label{eq:P5_e}  & \quad \text{Diag}(\mathbf{X}_n^{(i)}) = \mathbf{1}_M, \\
    \label{eq:P5_f}  & \quad \norm{\mathbf{G}\tilde{\mathbf{s}}_m^{(i)}}_{p\to\infty} \leq \gamma, ~ m\in\{1,\dots,M\}, \\
    \label{eq:P5_g} & \quad \frac{1}{\sqrt{MN}}\norm{\mathbf{S}^{(i)} - \mathbf{S}_0}_F \leq \delta,\\
    \label{eq:P5_h} & \quad \mathbf{X}_n^{(i)} \succcurlyeq \ZEROV,\\
    \label{eq:P5_i} & \quad \mathbf{Q}_n^{(i)} \succcurlyeq \ZEROV,\\
    \label{eq:P5_j} & \quad b_n^{(i)}\mathbf{I}_M - {\mathbf{V}_n^{(i-1)}}^{\dagger}\mathbf{Q}_n^{(i)}\mathbf{V}_n^{(i-1)} \succcurlyeq \ZEROV,\\
    \label{eq:P5_k} & \quad b_n^{(i-1)} \geq b_n^{(i)} \geq 0,
  \end{empheq}
\end{subequations}
\normalsize

Once $\mathbf{X}_n^{(i)}$, $\mathbf{S}_n^{(i)}$ and $b_n^{(i)}$ are found by solving \eqref{eq:P5}, $\mathbf{V}_n^{(i)}$ can be obtained by seeking an $(M+1) \times M$ matrix with orthonormal columns such that $b_n^{(i)}\mathbf{I}_M \succcurlyeq {\mathbf{V}_n^{(i)}}^{\dagger}\mathbf{Q}_n^{(i)}\mathbf{V}_n^{(i)}$. Choosing $\mathbf{V}_n^{(i)}$ to be equal to the matrix composed of the eigenvectors of $\mathbf{Q}_n^{(i)}$ corresponding to its $M$ smallest eigenvalues, and following similar arguments provided after \eqref{eq:P4} in the Appendix, we have \cite[Corollary 4.3.16]{horn2012matrix},

% \begin{equation}
%     {\mathbf{V}_n^{(i)}}^{\dagger} \mathbf{Q}_n^{(i)} \mathbf{V}_n^{(i)} = \text{Diag}([\rho_1^{(i)}, \rho_2^{(i)}, \cdots, \rho_M^{(i)}]^T) \nonumber\\
% \\preccurlyeq \text{Diag}([\nu_1^{(i-1)}, \nu_2^{(i-1)}, \cdots, \nu_M^{(i-1)}]^T) 
%  \preccurlyeq b_n^{(i)} \mathbf{I}_M,
% \end{equation}

\begin{equation}\label{eq:new-CH5-1}
\begin{aligned}
&{\mathbf{V}_n^{(i)}}^{\dagger} \mathbf{Q}_n^{(i)} \mathbf{V}_n^{(i)} = \text{Diag}([\rho_1^{(i)}, \rho_2^{(i)}, \cdots, \rho_M^{(i)}]^T)\\
&\preccurlyeq \text{Diag}([\nu_1^{(i-1)}, \nu_2^{(i-1)}, \cdots, \nu_M^{(i-1)}]^T) 
 \preccurlyeq b_n^{(i)} \mathbf{I}_M,
\end{aligned}
\end{equation}
where, $\rho_1^{(i)} \leq \rho_2^{(i)} \leq \cdots \leq \rho_{M+N}^{(i)}$ and $\nu_{1}^{(i-1)} \leq \nu_{2}^{(i-1)} \leq \cdots \leq \nu_M^{(i-1)}$ denote the eigenvalues of $\mathbf{Q}_n^{(i)}$ and ${\mathbf{V}_n^{(i-1)}}^{\dagger} \mathbf{Q}_n^{(i)} \mathbf{V}_n^{(i-1)}$, respectively. It follows from \eqref{eq:new-CH5-1} that the matrix composed of the eigenvectors of $\mathbf{Q}_n^{(i)}$ corresponding to its $M$ smallest eigenvalues is the appropriate choice for $\mathbf{V}_n^{(i)}$. 

Accordingly, at each iteration of the proposed algorithm which we term as \gls{WISE}, we need to solve a \gls{SDP} followed by an Eigenvalue Decomposition (EVD). \textbf{Algorithm \ref{alg:Alg1}} summarizes the steps of the \gls{WISE} approach for solving \eqref{eq:P0}. In order to initialize the algorithm, $\mathbf{V}_n^{(0)}$ can be found through the eigenvalue decomposition of $\mathbf{Q}_n^{(0)}$ obtained from solving \eqref{eq:P5} without considering \eqref{eq:P5_j} and \eqref{eq:P5_k} constraints. Further, we terminate the algorithm when
% $\mathbf{X}_n$ converges to a rank one solution and
$\bar{\mathbf{s}}_n\bar{\mathbf{s}}_n^{\dagger}$ converges to $\mathbf{X}_n$.
% , respectively. 
In this regards, let us assume that,
$$\xi_{n,1} \geq \xi_{n,2} \geq \dots \geq \xi_{n,m} \geq \dots \geq \xi_{n,M} \geq 0,$$ 
be the eigenvalues of $\mathbf{X}_n$, we consider $\xi \triangleq \frac{\max \{\xi_{n,2}\}}{\min \{\xi_{n,1}\}} < e_1$ ($e_1>0$) as the termination condition. In this case the {\it second} largest eigenvalue of $\mathbf{X}_n$ is negligible comparing to its largest eigenvalue and can be concluded that the solution is rank one. In addition, we consider $\max \left\{ \norm{\bar{\mathbf{s}}_n\bar{\mathbf{s}}_n^{\dagger} - \mathbf{X}_n}_F\right\} < e_2$ ($e_2>0$) as the second termination condition.

We note that the proposed algorithm, which is based on alternating optimization method, is guaranteed that the objective function converges to at least a local minimum of \eqref{eq:P3} \cite{bezdek2003convergence}. 

\begin{algorithm}[t]
	\caption{:\gls{WISE} in MIMO Radar Systems}	\label{alg:Alg1}
	\textbf{Input:} $\gamma$, $\delta$, $\mathbf{S}_0$, $\mathcal{U}$.\\
	\textbf{Initialization:} 
	\begin{enumerate}
	    \item $i := 0$;
	    \item Obtain $\mathbf{Q}_n^{(0)}$ by dropping \eqref{eq:P5_j} and \eqref{eq:P5_k} then solving \eqref{eq:P5};
	    \item $\mathbf{V}_n^{(0)}$ is the $M$ eigenvectors of $\mathbf{Q}_n^{(0)}$, corresponding to the $M$ lowest eigenvalues;
	    \item $b_n^{(0)}$ is the second largest eigenvalue of $\mathbf{Q}_n^{(0)}$;
	\end{enumerate}
	\textbf{Optimization:} 
	\begin{enumerate}
		\item {\bf while}, 
% 		$\sum_{n=1}^{N} \norm{\bar{\mathbf{s}}_n\bar{\mathbf{s}}_n^{\dagger} - \mathbf{X}_n}_F \geq e$,  
        % $\max \left\{ \norm{\bar{\mathbf{s}}_n\bar{\mathbf{s}}_n^{\dagger} - \mathbf{X}_n}_F\right\} \geq e$,  
        % $\xi = \frac{\min \{\xi_{n,1}\}}{\max \{\xi_{n,2}\}} < e$,
        $\xi \geq e_1$ and $\max \left\{ \norm{\bar{\mathbf{s}}_n\bar{\mathbf{s}}_n^{\dagger} - \mathbf{X}_n}_F\right\} \geq e_2$
		{\bf do}
		\item \hspace{5mm} $i := i+1$;
		\item \hspace{5mm} Obtain $\mathbf{S}^{(i)}$, $\mathbf{X}_n^{(i)}$ and $b_n^{(i)}$ by solving \eqref{eq:P5};
		\item \hspace{5mm} $\mathbf{V}_n^{(i)}$ is the $M$ eigenvectors of $\mathbf{Q}_n^{(i)}$, by dropping the eigenvector correspond to the largest eigenvalue.
	    \item \hspace{5mm} $b_n^{(i)}$ is the second largest eigenvalue of $\mathbf{Q}_n^{(i)}$.
		\item {\bf end while}
		\end{enumerate}
	\textbf{Output:} $\mathbf{S}^{\star} = \mathbf{S}^{(i)}$.
\end{algorithm}

\subsubsection{Convergence}\label{subsec:Convergence}
It readily follows from \eqref{eq:P5_k} that $\lim_{k \to \infty} \frac{|b_n^{(i)}|}{|b_n^{(i-1)}|} \leq 1$. This implies that $b_n^{(i)}$ converges at least sub-linearly to zero \cite{senning2007computing}. Hence, there exists some ${\cal{I}}$ such that $b_n^{(i)} \leq \epsilon$ ($\epsilon \to 0$) for $i \geq {\cal{I}}$. Making use of this fact, we can deduce from \eqref{eq:P5_j} that,
\begin{align}
\label{eq:conv-1}
{\mathbf{V}_n^{(i-1)}}^{\dagger}\mathbf{Q}_n^{(i)}\mathbf{V}_n^{(i-1)} \preccurlyeq \epsilon \mathbf{I}_M, \quad \epsilon \to 0,
\end{align}
for $i \geq {\cal{I}}$. Then, it follows from \eqref{eq:conv-1} and \eqref{eq:new-CH5-1} that $\text{Rank}(\mathbf{Q}_n^{(i)}) \simeq 1$ for $i \geq {\cal{I}}$, thereby $\mathbf{X}_n^{(i)} = \bar{\mathbf{s}}_n^{(i)}\bar{\mathbf{s}}_n^{(i)\dagger}$ for $i \geq {\cal{I}}$. This implies that $\mathbf{X}_n^{(i)}$, for any $i \geq {\cal{I}}$, is a feasible point for the optimization problem \eqref{eq:P3}. On the other hand, considering the fact that $b_n^{(i)} \leq \epsilon$ for $i \geq {\cal{I}}$, we conclude that $\mathbf{X}_n^{(i)}$ for $i \geq {\cal{I}}$ is also a minimizer of the function $\sum_{n=1}^{N}\text{Tr}(\mathbf{A}_u\mathbf{X}_n)$. These imply that $\mathbf{X}_n^{(i)}$ for $i \geq {\cal{I}}$ is at least a local minimizer of the optimization problem \eqref{eq:P3}. The proves the convergence of the proposed iterative algorithm.

\subsubsection{Computational Complexity}\label{subsec:Complexity}
In each iteration, \textbf{Algorithm \ref{alg:Alg1}} needs to perform the following steps:
\begin{itemize}
    \item {\it Solving \eqref{eq:P5}}: Needs the solution of a \gls{SDP}, whose computational complexity is ${\cal{O}}(M^{3.5})$ \cite{5447068}.
    \item {\it Obtaining $\mathbf{V}_n^{(i)}$ and $b_n^{(i)}$}: Needs the implementation of a \gls{SVD}, whose computational complexity is ${\cal{O}}(M^3)$ \cite{GoluVanl96}.
\end{itemize}
Let us assume that ${\cal{I}}$ iterations are required for convergence of the \textbf{Algorithm \ref{alg:Alg1}}. Therefore, the overall computational complexity of \textbf{Algorithm \ref{alg:Alg1}} is, ${\cal{O}}({\cal{I}}(M^{3.5} + M^3))$.

\section{Numerical Results}\label{sec:Numerical Results}
In this section, numerical results are provided for assessing the performance of the proposed algorithm for beampattern shaping and spectral matching under constant modulus constraint. Towards this end, unless otherwise explicitly stated, we consider the following set-up. For transmit parameters, we consider \gls{ULA} configuration with $M=8$ transmitters, with the spacing of $d=\lambda/2$ and each antenna transmits $N=64$ samples.
% each transmits $N=64$ samples and the antenna distance is set as $d = \frac{\lambda}{2}$. 
We consider an uniform sampling of regions $\theta = [-90^o, 90^o]$ with a grid size of $5^o$ and the desired and undesired angels for beampattern shaping are $\Theta_d = [-55^o, -35^o]$ ($\theta_0 = -45^o$) and $\Theta_u = [-90^o, -60^o] \cup [-30^o, 90^o]$, respectively. The normalized frequency stop-band is set at $\mathcal{U} = [0.3, 0.35] \cup [0.4, 0.45] \cup[0.7, 0.8]$ and the absolute spectral mask level is set as $\gamma = 0.01\sqrt{N}$. As to the reference signal for similarity constraint, we consider $\mathbf{S}_0$ be a set of sequences with a good range-\gls{ISLR} property, which is obtained by the method in \cite{9440807}. For the optimization problem we set $\eta=0.1$ and $p=1000$ to approximate the \eqref{eq:P1_f} constraint. The convex optimization problems are solved via the CVX toolbox \cite{CVX_web} and the stop condition for \textbf{Algorithm~\ref{alg:Alg1}} are set at $e_1=10^{-5}$ and $e_2=10^{-4}$, respectively.  

\subsection{Convergence} 
\begin{figure}
    \centering
    \subfloat[Convergence of eigenvalue]{\includegraphics[width=1\columnwidth]{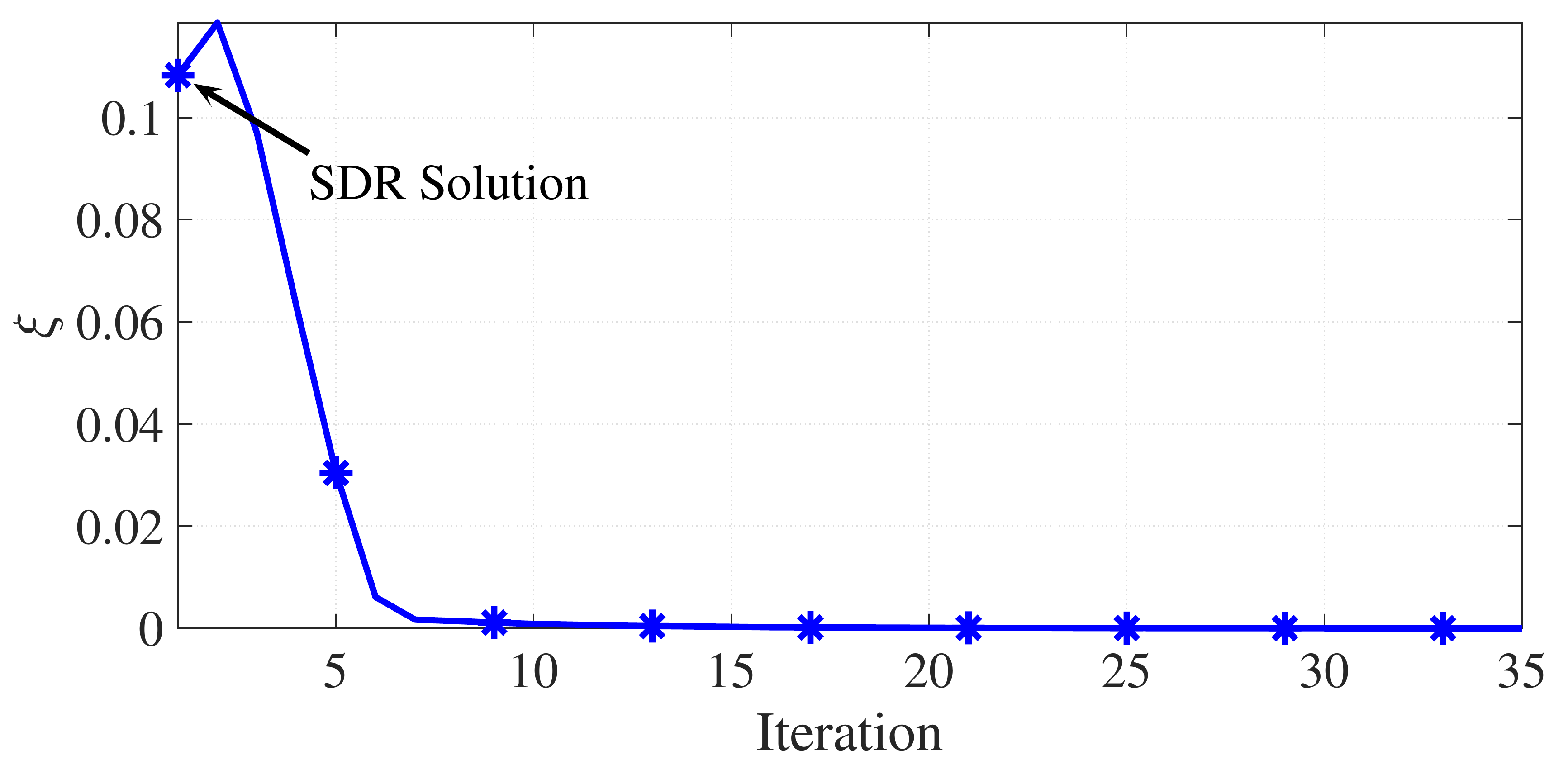}%
    \label{fig:Convergence_EigenValue}}
    \\
    \subfloat[Convergence of arguments]{\includegraphics[width=1\columnwidth]{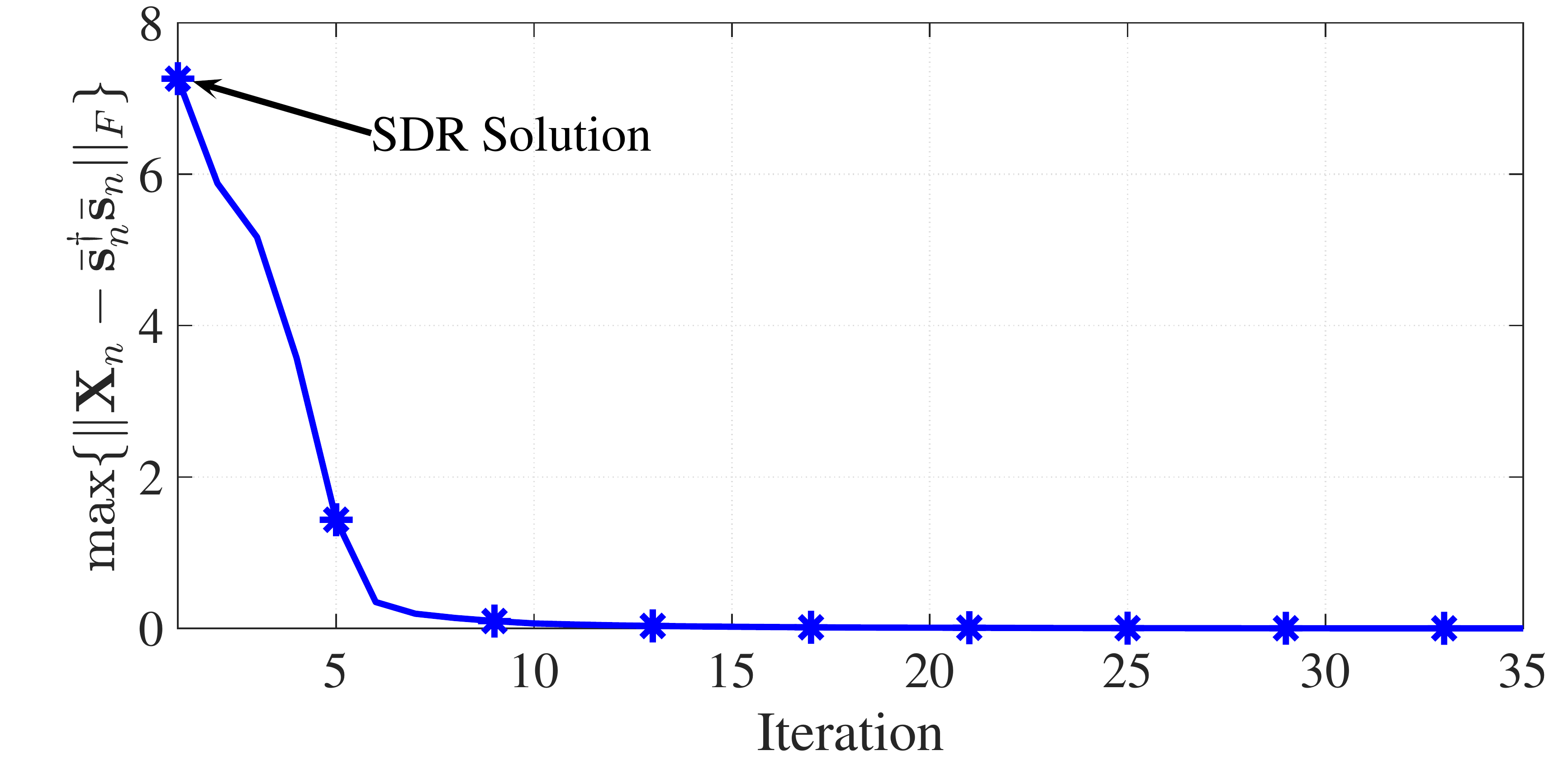}%
    \label{fig:Convergence_dSX}}
    \\
    \subfloat[Convergence of constant modulus]{\includegraphics[width=1\columnwidth]{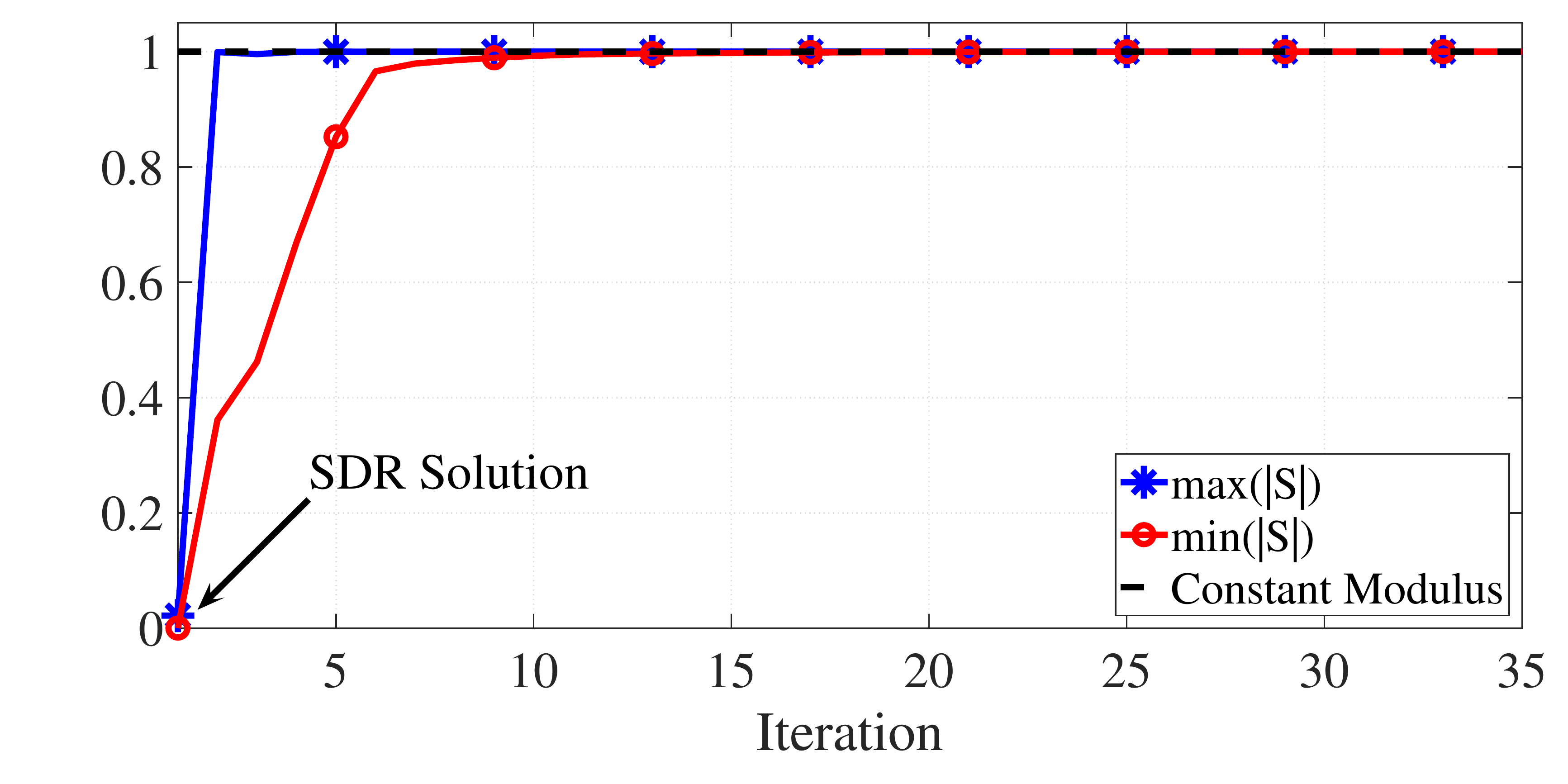}%
    \label{fig:Convergence_ConstantModulus}}
    \caption{The convergence behavior of proposed method in different aspects, (a) $\xi = \frac{\max \{\xi_{n,2}\}}{\min \{\xi_{n,1}\}}$, (b) $\max \left\{ \norm{\mathbf{X}_n - \mathbf{s}_n^{\dagger}\mathbf{s}_n}_F\right\}$  and (c) Constant modulus ($M=8$, $N=64$, $\delta = \sqrt{2}$, $\Theta_d = [-55^o, -35^o]$, $\Theta_u = [-90^o, -60^o] \cup [-30^o, 90^o]$, $\mathcal{U} = [0.3, 0.4]$ and $\gamma = 0.01\sqrt{N}$}
    \label{fig:Convergence}
\end{figure}

\figurename{~\ref{fig:Convergence}} depicts the convergence behavior of the proposed method in different aspects. In this figure, we consider the maximum admissible value for similarity parameter, i.e., $\delta = \sqrt{2}$. \figurename{~\ref{fig:Convergence_EigenValue}} shows the convergence of $\xi$ to zero. This indicates that the second largest eigenvalue of $\mathbf{X}_n$ is negligible in comparison with the largest eigenvalue, therefore resulting in a rank one solution for $\bar{\mathbf{s}}_n$. \figurename{~\ref{fig:Convergence_dSX}} shows that the solution of $\mathbf{X}_n$ converges to $\bar{\mathbf{s}}_n$, which confirms our claim about obtaining a rank one solution. \figurename{~\ref{fig:Convergence_ConstantModulus}} shows the convergence of the $\max\{|\mathbf{S}|\}$ to $\min\{|\mathbf{S}|\}$, which indicates the constant modulus solution.

Please note that, the first iteration in \figurename{~\ref{fig:Convergence_EigenValue}}, \figurename{~\ref{fig:Convergence_dSX}} and \figurename{~\ref{fig:Convergence_ConstantModulus}} shows the \gls{SDR} solution of \eqref{eq:P5} by dropping \eqref{eq:P5_j} and \eqref{eq:P5_k}. As can be seen the \gls{SDR} method offers neither rank one nor constant modulus solution.

% \begin{figure*}[h]
%     \centering
%     \begin{subfigure}{.49\textwidth}
%         \centering
% 		\includegraphics[width=1\linewidth]{figures/Convergence_EigenValue_M8_N64.eps}
% 		\caption[]{} \label{fig:Convergence_EigenValue}
%     \end{subfigure}
%     \begin{subfigure}{.49\textwidth}
%         \centering
% 		\includegraphics[width=1\linewidth]{figures/Convergence_dSX_M8_N64.eps}
% 		\caption[]{} \label{fig:Convergence_dSX}
%     \end{subfigure}
%     \begin{subfigure}{.49\textwidth}
%         \centering
% 		\includegraphics[width=1\linewidth]{figures/Convergence_ConstantModulus_M8_N64.eps}
% 		\caption[]{} \label{fig:Convergence_ConstantModulus}
%     \end{subfigure}
%     \caption[]{The convergence behavior of proposed method in different aspects, (a) $\xi = \frac{\max \{\xi_{n,2}\}}{\min \{\xi_{n,1}\}}$, (b) Constant modulus and (c) $\max \left\{ \norm{\mathbf{X}_n - \mathbf{s}_n^{\dagger}\mathbf{s}_n}_F\right\}$ ($M=8$, $N=64$, $\delta = \sqrt{2}$, $\Theta_d = [-55^o, -35^o]$, $\Theta_u = [-90^o, -60^o] \cup [-30^o, 90^o]$, $\mathcal{U} = [0.3, 0.4]$ and $\gamma = 0.01\sqrt{N}$).}\label{fig:Convergence}
% \end{figure*}

% \begin{figure*}[h]
%     \centering
%     \begin{subfigure}{.32\textwidth}
%         \centering
% 		\includegraphics[width=1\linewidth]{figures/Convergence_EigenValue_M8_N64.eps}
% 		\caption[]{} \label{fig:Convergence_EigenValue}
%     \end{subfigure}
%     \begin{subfigure}{.32\textwidth}
%         \centering
% 		\includegraphics[width=1\linewidth]{figures/Convergence_dSX_M8_N64.eps}
% 		\caption[]{} \label{fig:Convergence_dSX}
%     \end{subfigure}
%     \begin{subfigure}{.32\textwidth}
%         \centering
% 		\includegraphics[width=1\linewidth]{figures/Convergence_ConstantModulus_M8_N64.eps}
% 		\caption[]{} \label{fig:Convergence_ConstantModulus}
%     \end{subfigure}
%     \caption[]{The convergence behavior of proposed method in different aspects, (a) $\xi = \frac{\max \{\xi_{n,2}\}}{\min \{\xi_{n,1}\}}$, (b) Constant modulus and (c) $\max \left\{ \norm{\mathbf{X}_n - \mathbf{s}_n^{\dagger}\mathbf{s}_n}_F\right\}$ ($M=8$, $N=64$, $\delta = \sqrt{2}$, $\Theta_d = [-55^o, -35^o]$, $\Theta_u = [-90^o, -60^o] \cup [-30^o, 90^o]$, $\mathcal{U} = [0.3, 0.4]$ and $\gamma = 0.01\sqrt{N}$).}\label{fig:Convergence}
% \end{figure*}

\subsection{Performance} \label{subsec:Beampattern Shaping and Spectral Masking}
\begin{figure*}%[h]
    \centering
    \begin{subfigure}{.49\textwidth}
        \centering
		\includegraphics[width=1\linewidth]{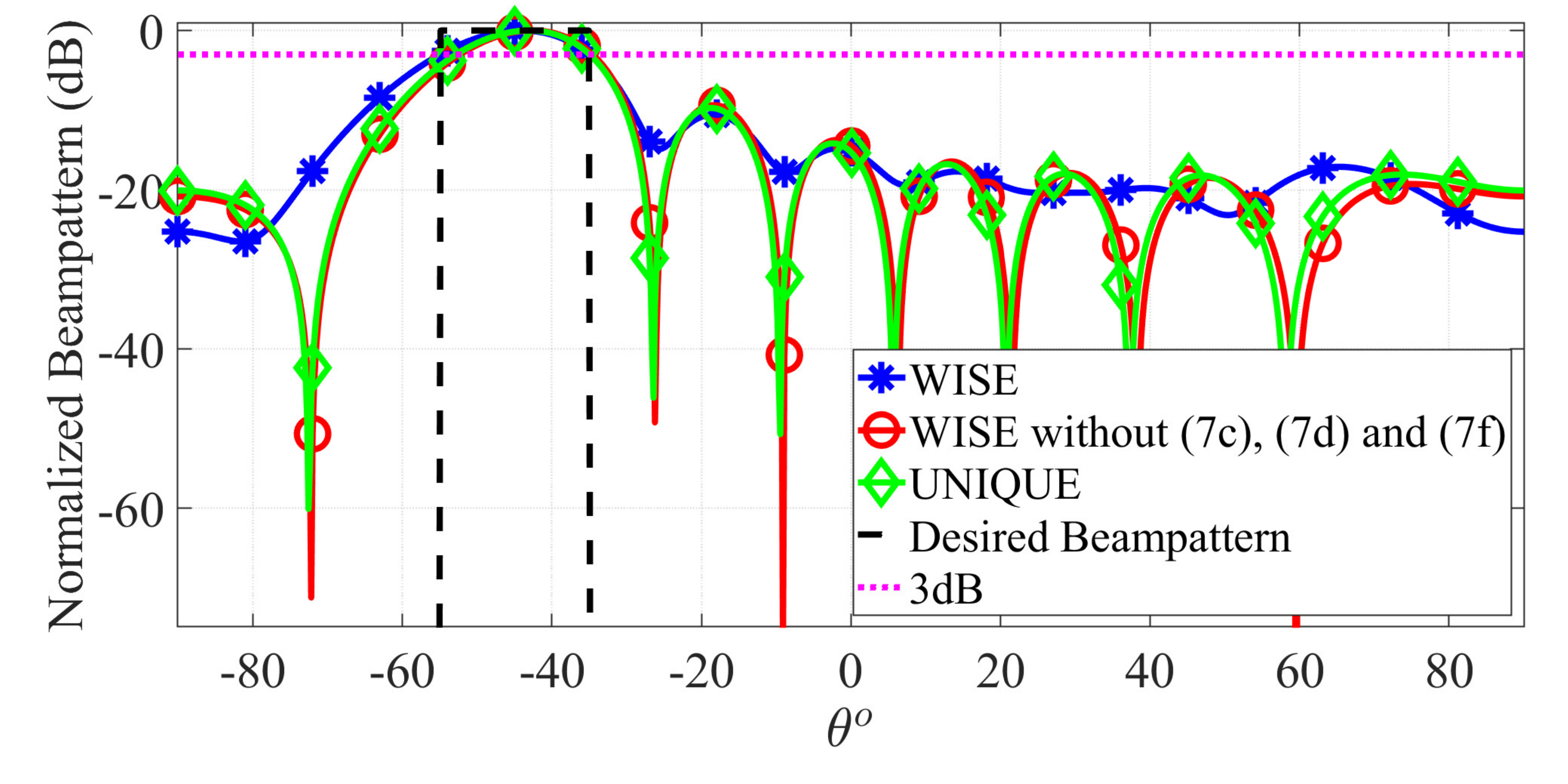}
		\caption[]{Beampattern response} \label{fig:BeamPattern WISE vs UNIQUE}
    \end{subfigure}
    \begin{subfigure}{.49\textwidth}
        \centering
		\includegraphics[width=1\linewidth]{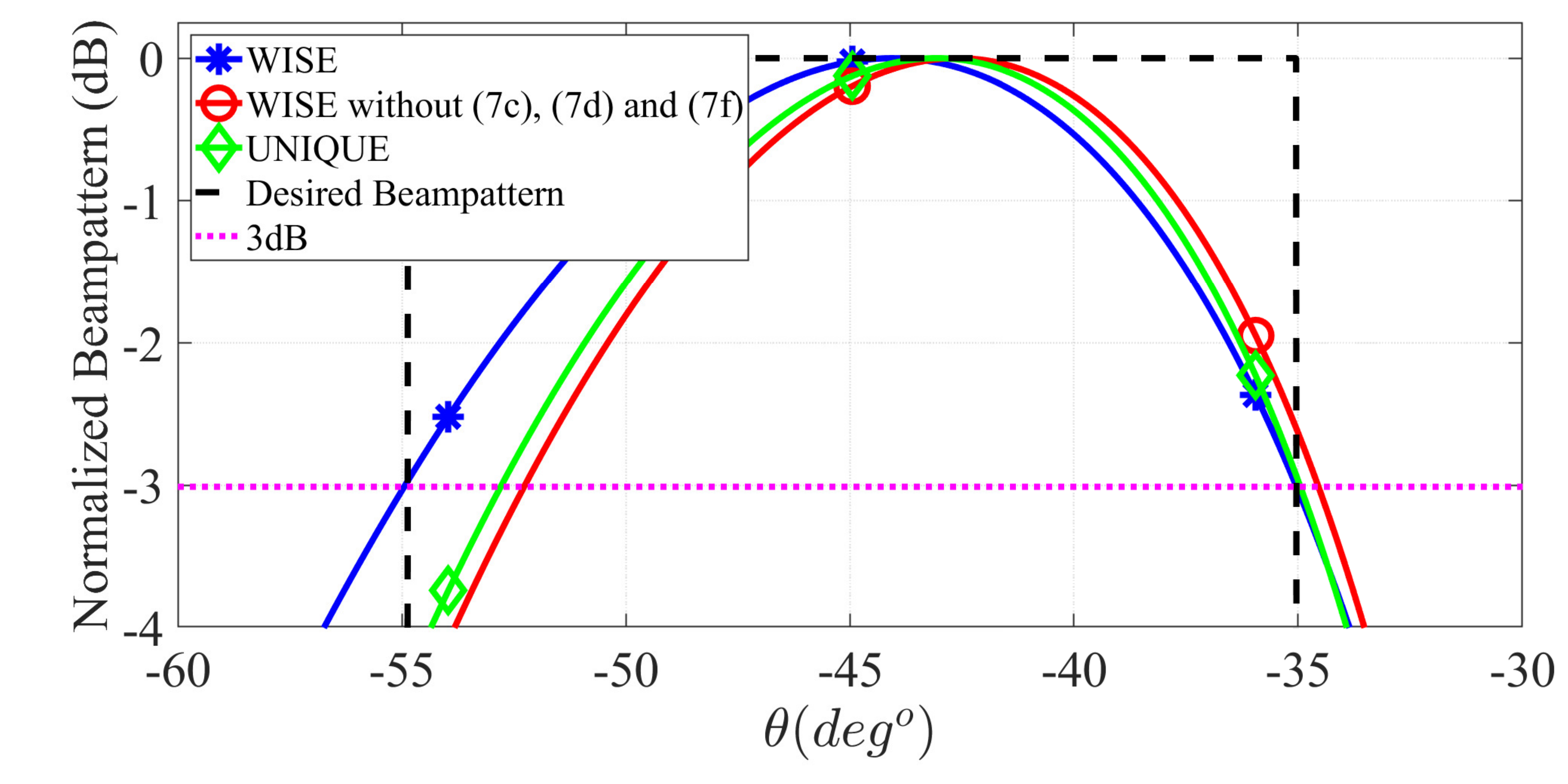}
		\caption[]{$3$ dB main beam-width} \label{fig:BeamPattern_3dB}
    \end{subfigure}
    \begin{subfigure}{.49\textwidth}
        \centering
		\includegraphics[width=1\linewidth]{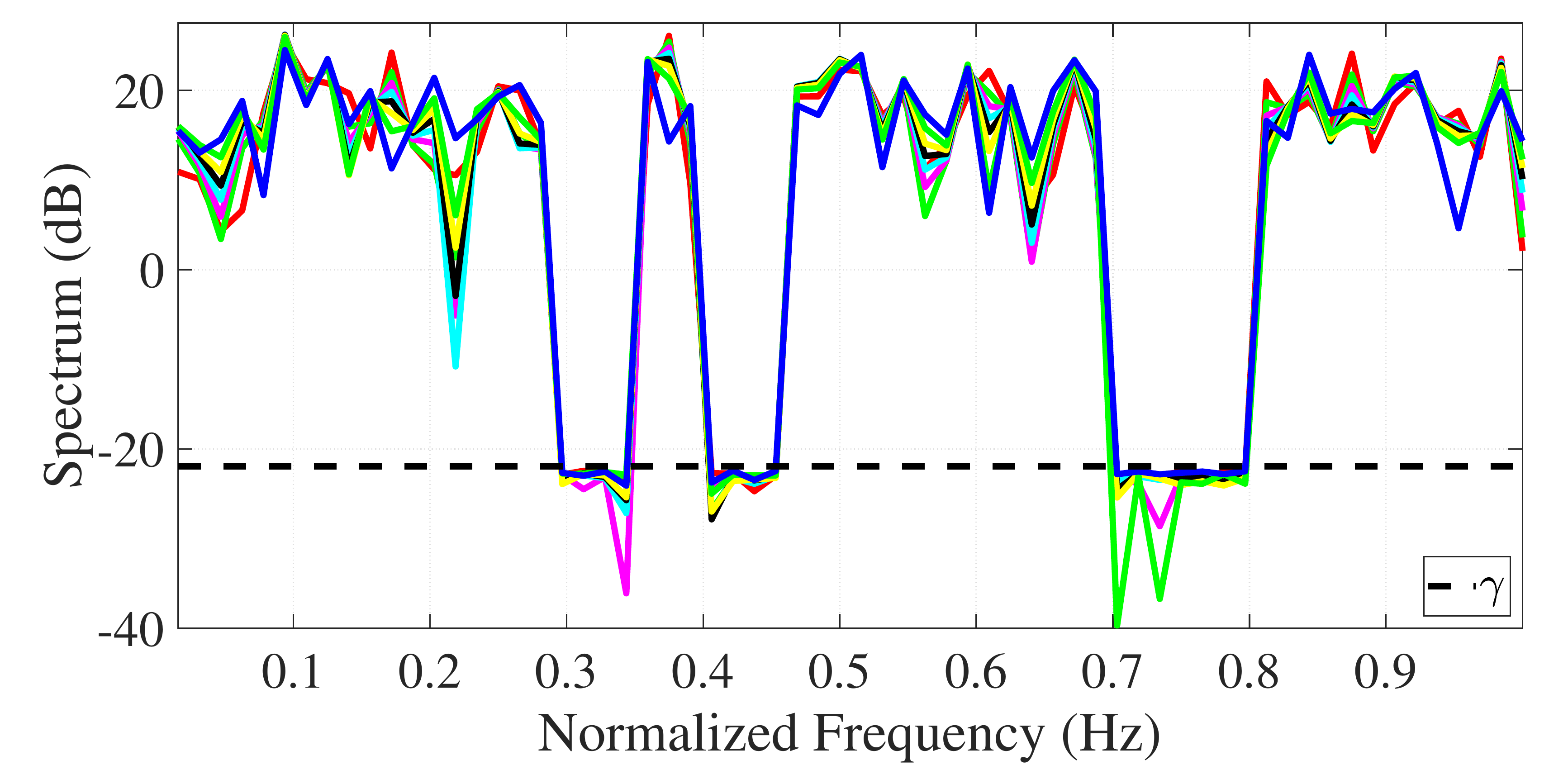}
		\caption[]{Spectral of \gls{WISE}} \label{fig:SpectralMasking_WISE}
    \end{subfigure}
    \begin{subfigure}{.49\textwidth}
        \centering
		\includegraphics[width=1\linewidth]{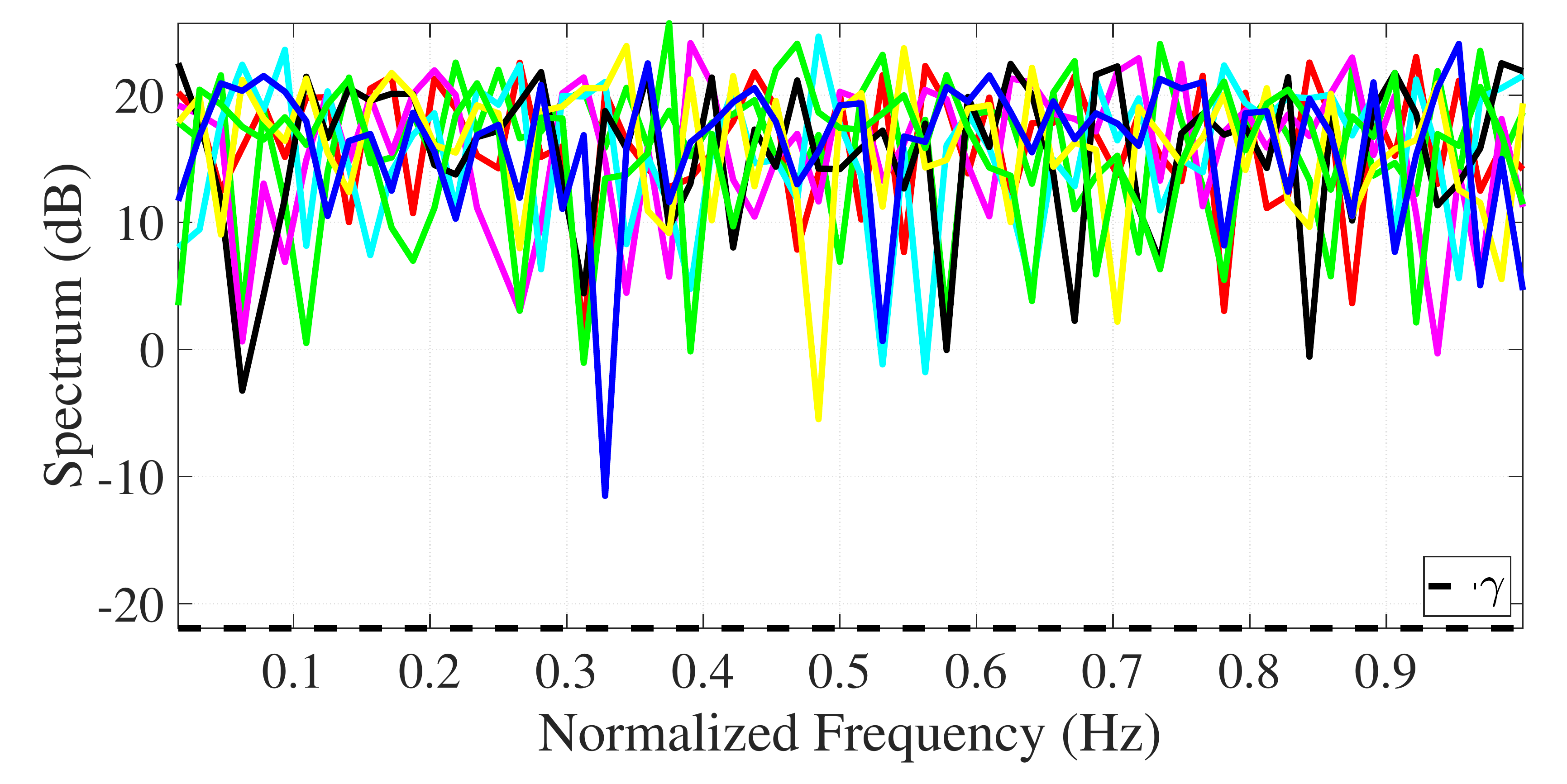}
		\caption[]{Spectral of \gls{UNIQUE}} \label{fig:SpectralMasking_UNIQUE}
    \end{subfigure}
    \caption[]{Comparing the performance of \gls{WISE} and \gls{UNIQUE} methods in terms of (a) $3$ dB main beam-width constraint, (b) spectral masking of \gls{WISE} and (c) spectral masking of \gls{UNIQUE} ($M=4$, $N=64$, $\delta = \sqrt{2}$, $\Theta_d = [-55^o, -35^o]$, $\Theta_u = [-90^o, -60^o] \cup [-30^o, 90^o]$, $\mathcal{U} = [0.3, 0.35] \cup [0.4, 0.55] \cup [0.7, 0.85]$ and $\gamma = 0.01\sqrt{N}$).}\label{fig:BP3dBSpectral}
\end{figure*}

\figurename{~\ref{fig:BP3dBSpectral}} compares the performance of the proposed method in terms of beampattern shaping and spectral masking with \gls{UNIQUE} \cite{9440807} method as a benchmark. \figurename{~\ref{fig:BeamPattern WISE vs UNIQUE}} shows the beampattern response of the proposed method and \gls{UNIQUE}. In this figure, for fair comparison we drop the spectral masking \eqref{eq:P5_f} and $3$ dB main beam-width \eqref{eq:P5_c} and \eqref{eq:P5_d} constraints. As can be seen, in this case the proposed method offers almost similar performance (in some undesired angles deeper nulls) as compared to \gls{UNIQUE} method. However, considering the spectral masking \eqref{eq:P1_f} and $3$ dB main beam-width \eqref{eq:P5_c} and \eqref{eq:P5_d} constraints, the proposed method 
% mimics the beampattern of \gls{UNIQUE} method and 
is able to steer the beam towards the desired and steer the nulls at undesired angles simultaneously. 

% \begin{figure}[h]
%     \centering
%         \centering
% 		\includegraphics[width=1\linewidth]{figures/BP1_WISE_vs_UNIQUE.eps}
% 		\caption[]{Comparing the beampattern response of \gls{WISE} and \gls{UNIQUE} method ($M=8$, $N=64$, $\delta = \sqrt{2}$, $\Theta_d = [-55^o, -35^o]$ and $\Theta_u = [-90^o, -60^o] \cup [-30^o, 90^o]$, $\mathcal{U} = [0.3, 0.4]$ and $\gamma = 0.01\sqrt{N}$).} \label{fig:BeamPattern WISE vs UNIQUE}
% \end{figure}

% Although the \gls{UNIQUE} method offers deeper nulls in compare with \gls{WISE} method, however, 
The beampattern response of \gls{WISE} at desired angles region and the spectrum response of the proposed method has better performance comparing to \gls{UNIQUE} method. \figurename{~\ref{fig:BeamPattern_3dB}} shows the main beam-width response of the proposed method and \gls{UNIQUE}. Since \gls{UNIQUE} does not have the $3$ dB main beam-width constraint, it method does not have a good main beam-width response. However, the $3$ dB main beam-width constraint incorporated in our framework improves the main beam-width response. Besides, the maximum beampattern response is located at $-45^o$ in the proposed method while there is a deviation in \gls{UNIQUE} method. On the other hand \figurename{~\ref{fig:SpectralMasking_WISE}} shows the spectrum response of the proposed method. Observe that the waveform obtained by \gls{WISE} masks the the spectral power in the stop-bands region ($\mathcal{U}$) below the $\gamma$ value. However, since \gls{UNIQUE} method is not spectral compatible, is unable to put notches on the stop-bands.

\subsection{The impact of similarity parameter} \label{The impact of delta}
In this subsection, we evaluate the impact of choosing the similarity parameter $\delta$ on performance of the proposed method. When we consider the maximum admissible value for similarity parameter, i.e., $\delta = \sqrt{2}$, we do not include similarity constraint and by decreasing $\delta$ we have the degree of freedom to enforce 
% the optimum solution have similar 
properties similar to the reference waveform on the optimal waveform. As mentioned in section \ref{sec:Numerical Results}, we consider $\mathbf{S}_0$ be a set of sequences with a good range-\gls{ISLR} property as the reference signal for similarity constraint, which is obtained by \gls{UNIQUE} method \cite{9440807}. Therefore, by decreasing the $\delta$ we obtain a waveform with good orthogonality, which leads to omni directional beampattern. 

\begin{figure}%[h]
    \centering
        \centering
		\includegraphics[width=1\linewidth]{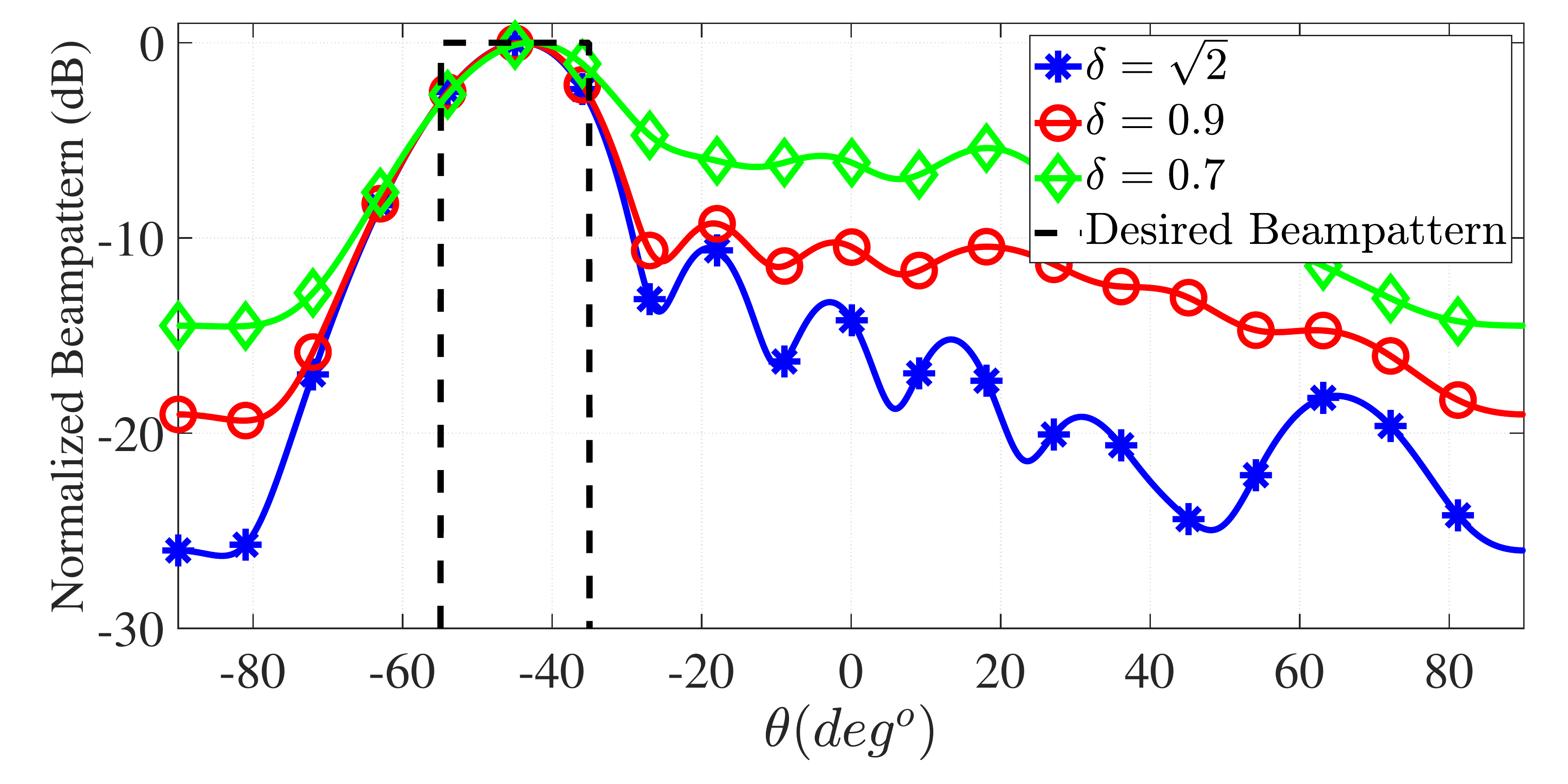}
		\caption[]{The impact of choosing $\delta$ in the proposed method on Beampattern response ($M=8$, $N=64$, $\Theta_d = [-55^o, -35^o]$ and $\Theta_u = [-90^o, -60^o] \cup [-30^o, 90^o]$, $\mathcal{U} = [0.3, 0.35] \cup [0.5, 0.55]$ and $\gamma = 0.01\sqrt{N}$).} \label{fig:BeamPattern WISE vs delta}
\end{figure}

\figurename{~\ref{fig:BeamPattern WISE vs delta}} shows the beampattern response of the proposed method with different values of $\delta$. As can be seen, with $\delta = \sqrt{2}$, yields an optimized beampattern and by decreasing $\delta$ the beampattern gradually tends to be omnidirectional. 

\begin{figure*}%[h]
    \centering
    \begin{subfigure}{.49\textwidth}
        \centering
		\includegraphics[width=1\linewidth]{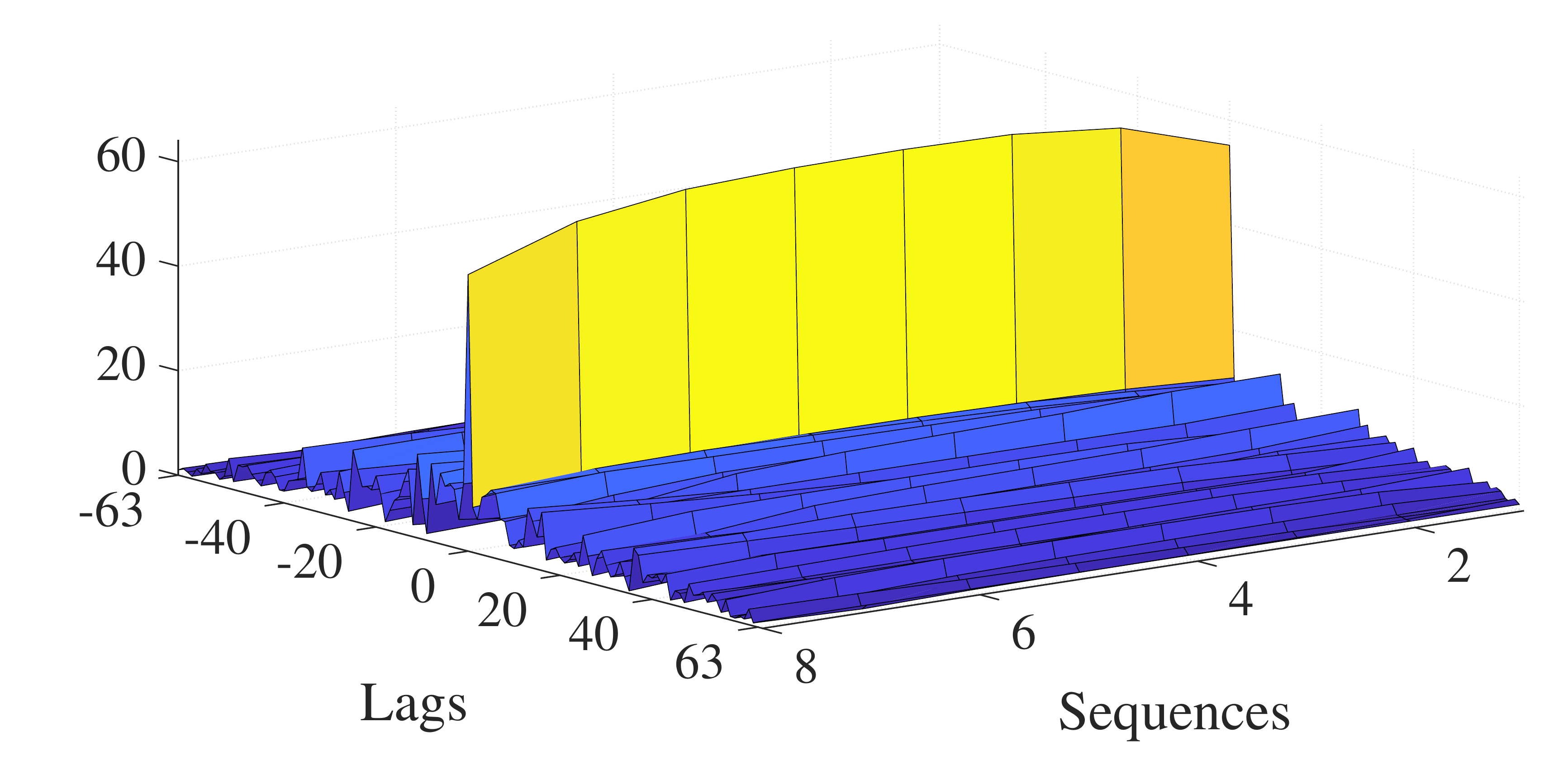}
		\caption[]{$\delta = \sqrt{2}$} \label{fig:CorrelationLevel_delta_sqrt2}
    \end{subfigure}
    \begin{subfigure}{.49\textwidth}
        \centering
		\includegraphics[width=1\linewidth]{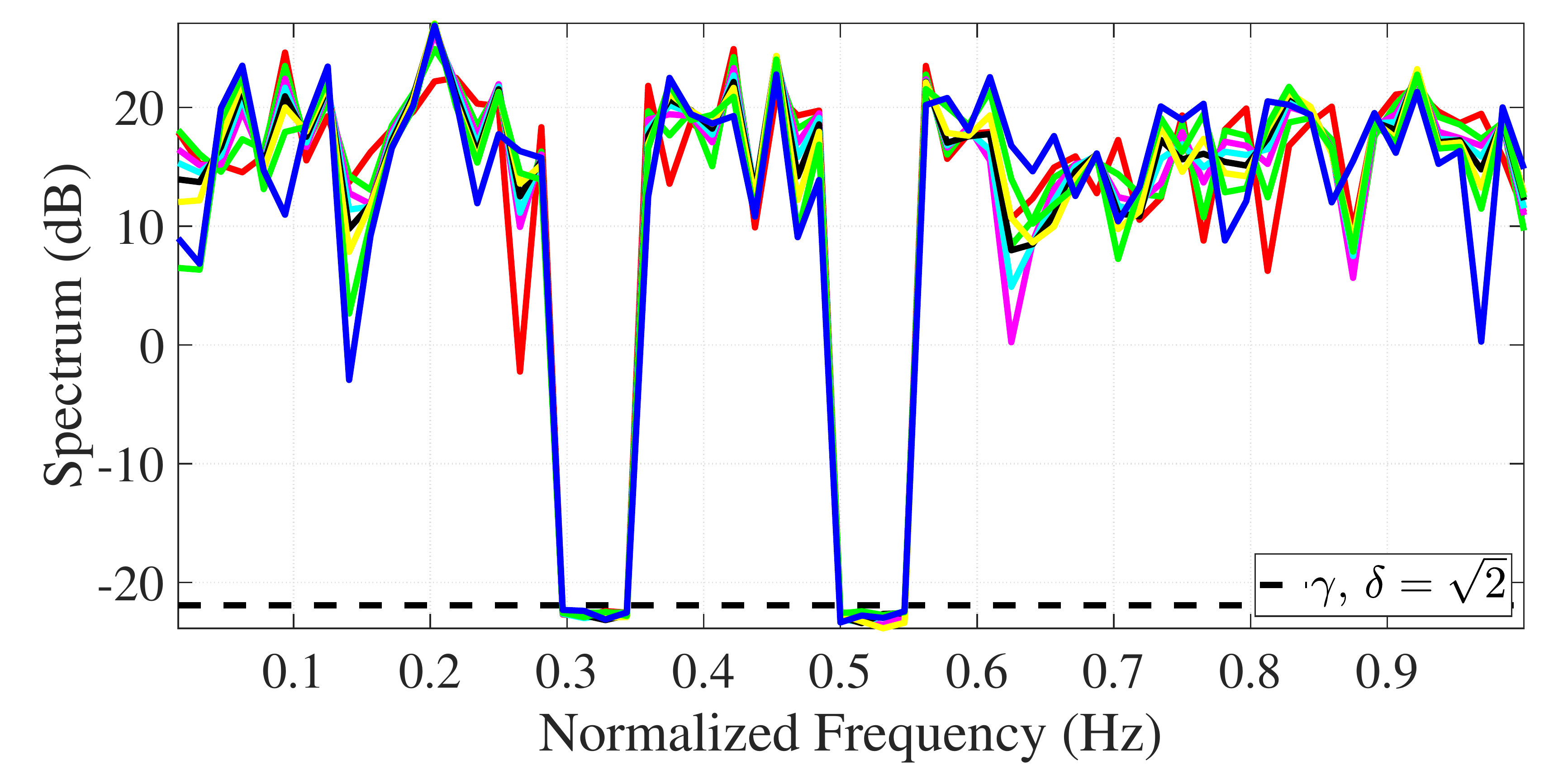}
		\caption[]{$\delta = \sqrt{2}$} \label{fig:SpectralMatching_delta_sqrt2}
    \end{subfigure}
    \begin{subfigure}{.49\textwidth}
        \centering
		\includegraphics[width=1\linewidth]{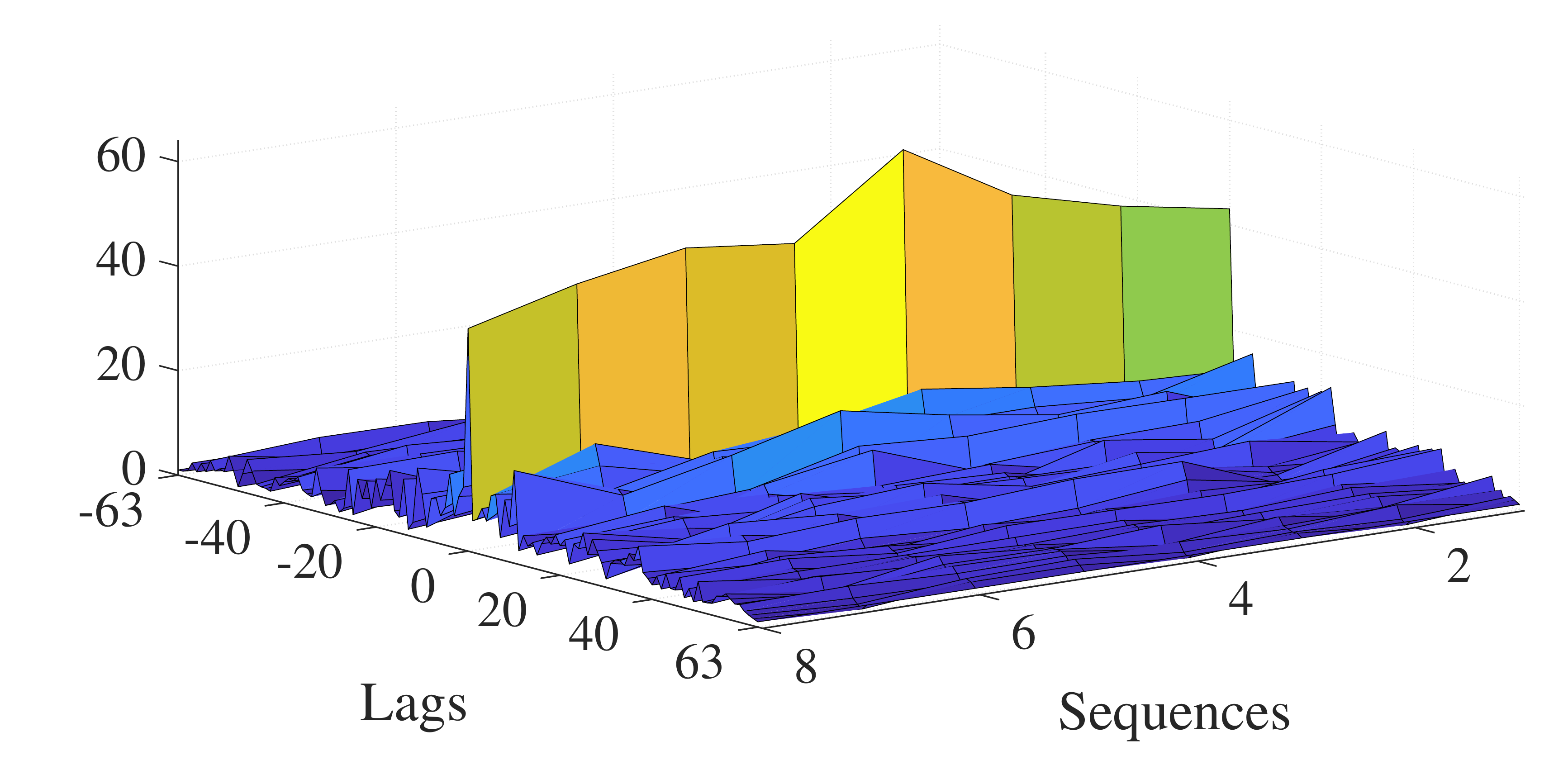}
		\caption[]{$\delta = 0.9$} \label{fig:CorrelationLevel_delta_09}
    \end{subfigure}
    \begin{subfigure}{.49\textwidth}
        \centering
		\includegraphics[width=1\linewidth]{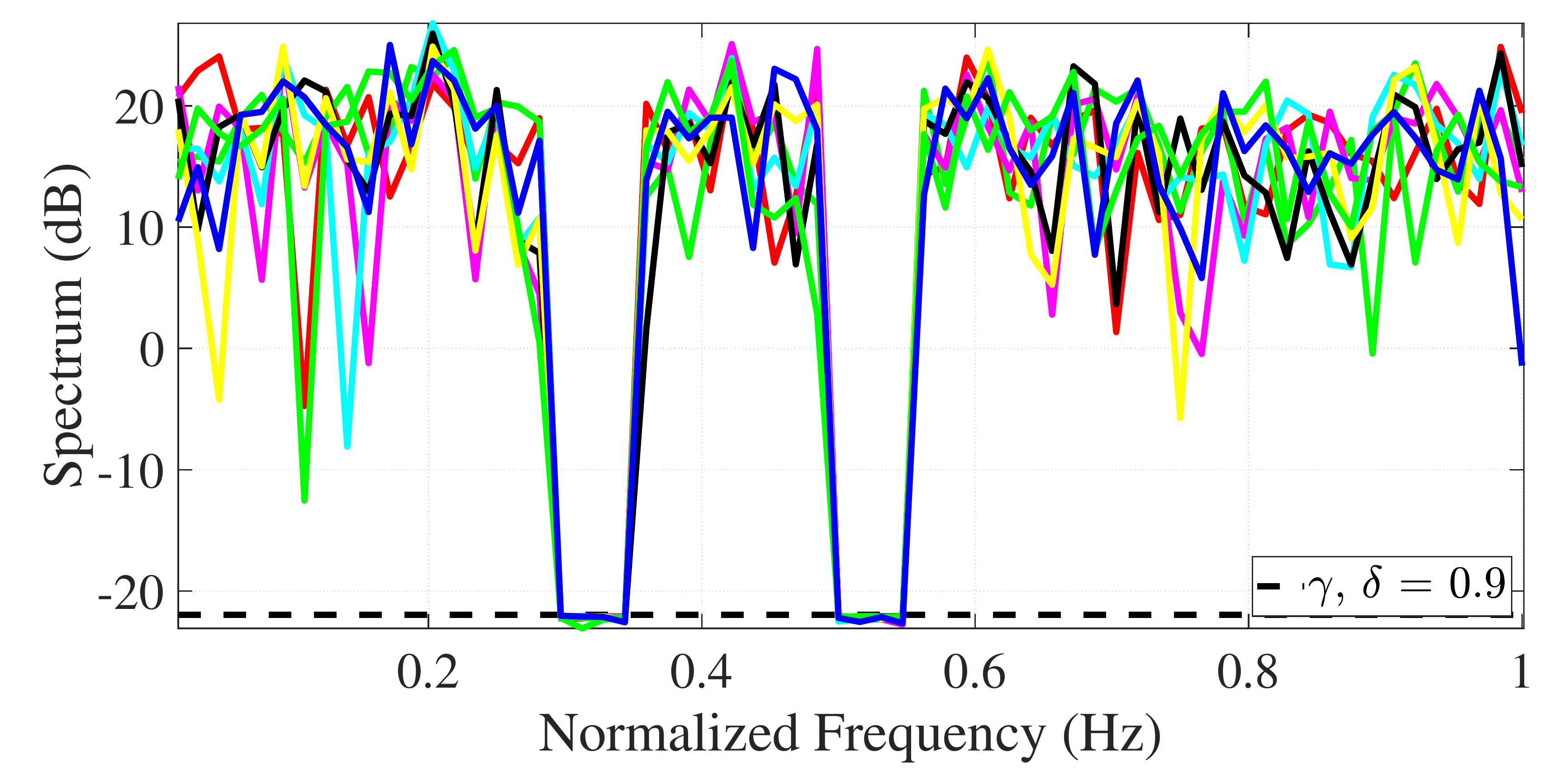}
		\caption[]{$\delta = 0.9$} \label{fig:SpectralMatching_delta_09}
    \end{subfigure}
    \begin{subfigure}{.49\textwidth}
        \centering
		\includegraphics[width=1\linewidth]{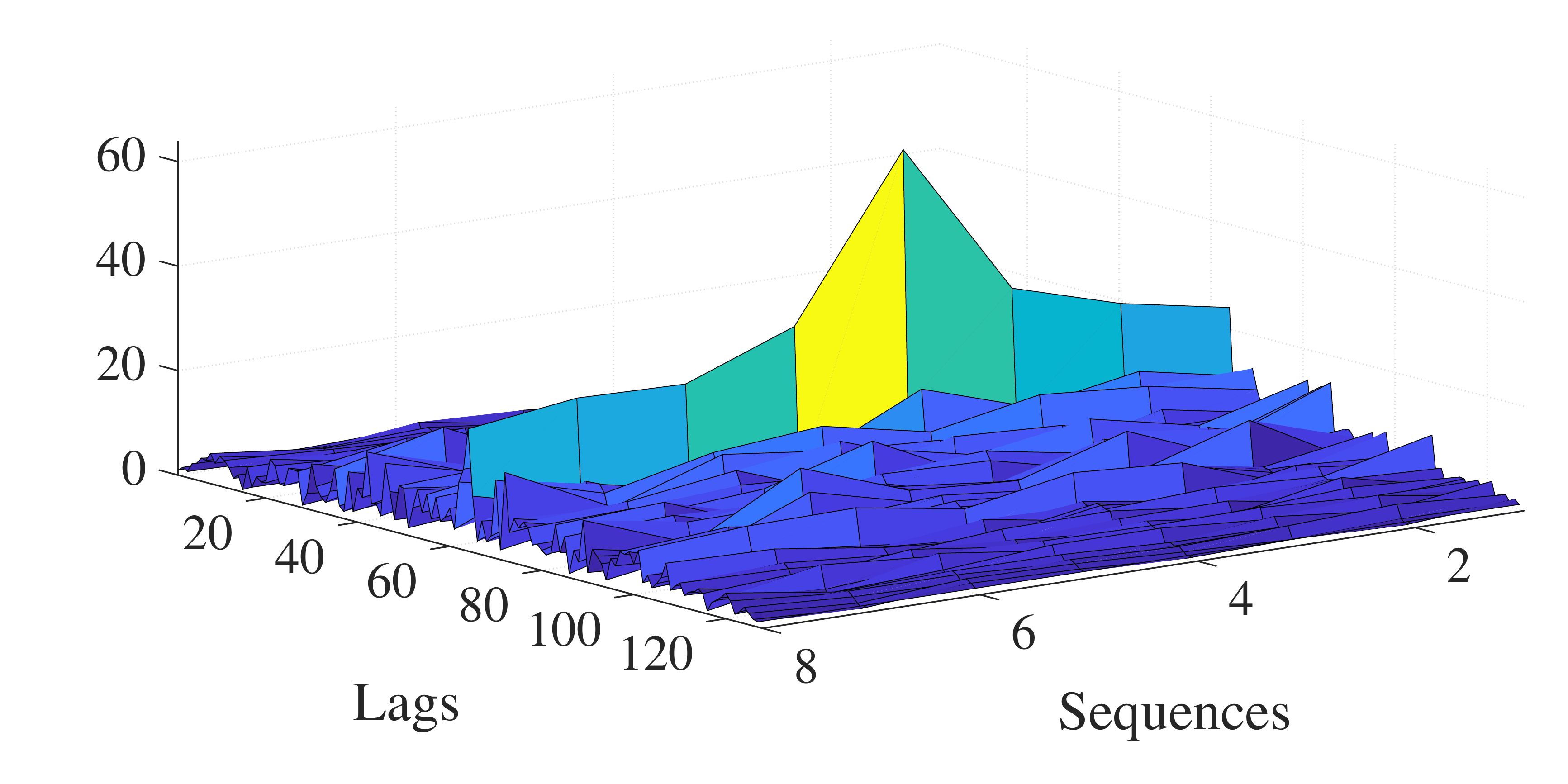}
		\caption[]{$\delta = 0.7$} \label{fig:CorrelationLevel_delta_07}
    \end{subfigure}
    \begin{subfigure}{.49\textwidth}
        \centering
		\includegraphics[width=1\linewidth]{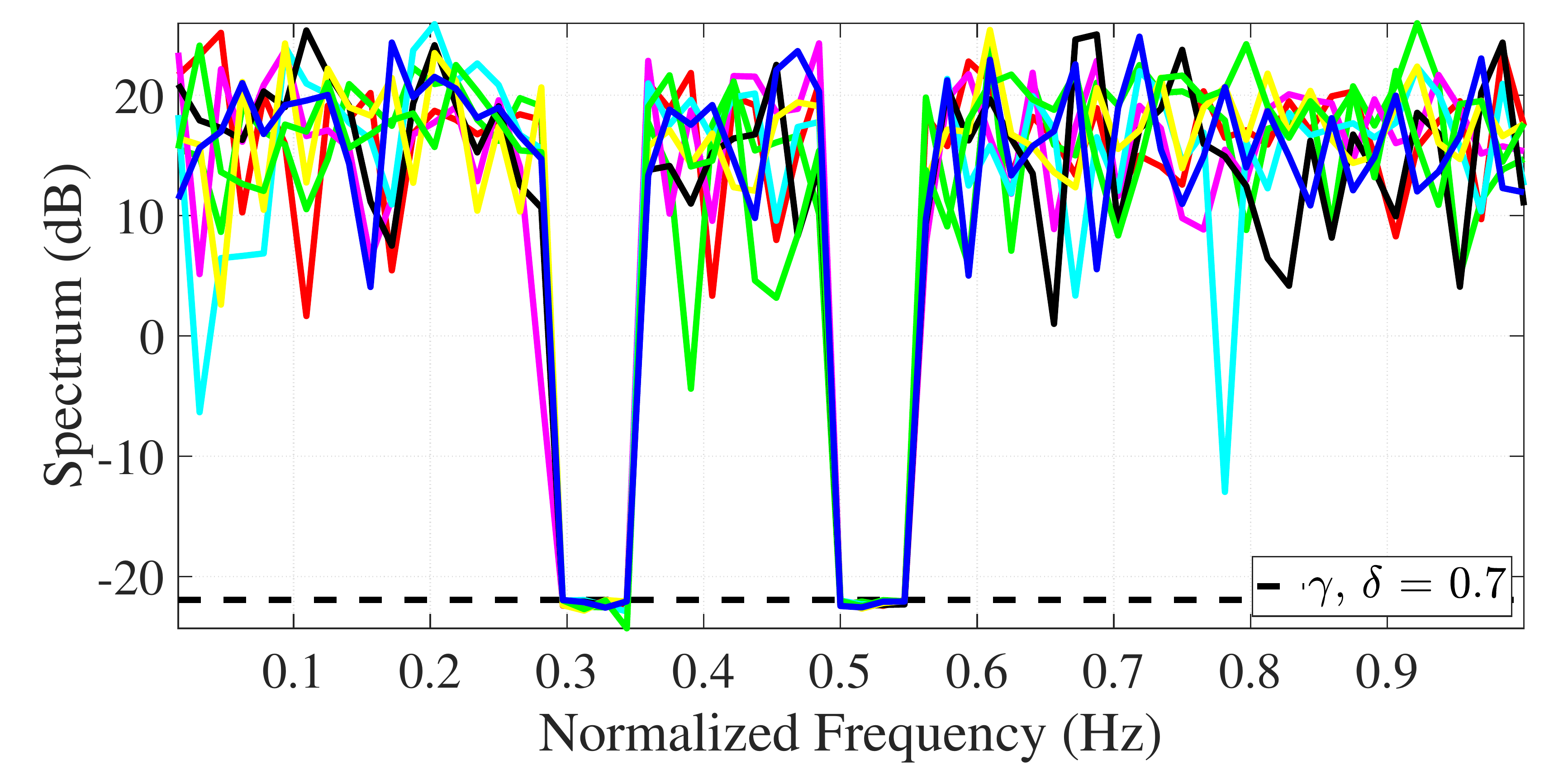}
		\caption[]{$\delta = 0.7$} \label{fig:SpectralMatching_delta_07}
    \end{subfigure}
    \caption[]{The impact of choosing $\delta$ on correlation level and spectral masking ($M=8$, $N=64$, $\Theta_d = [-55^o, -35^o]$, $\Theta_u = [-90^o, -60^o] \cup [-30^o, 90^o]$, $\mathcal{U} = [0.3, 0.35] \cup [0.5, 0.55]$ and $\gamma = 0.01\sqrt{N}$).}\label{fig:SM_CC_vs_delta}
\end{figure*}

On the other hand, \figurename{~\ref{fig:CorrelationLevel_delta_sqrt2}}, \figurename{~\ref{fig:CorrelationLevel_delta_09}} and \figurename{~\ref{fig:CorrelationLevel_delta_07}} show the correlation level of the proposed method with different values of $\delta$. Observe that with $\delta = \sqrt{2}$ we obtain fully correlated waveform and by decreasing $\delta$ the waveform gradually becomes uncorrelated. Therefore, having simultaneous beampattern shaping and orthogonality are contradictory, and the choice of $\delta$ effects a trade-off between the  two and enhance the performance of radar system \cite{9440807}. Besides \figurename{~\ref{fig:SpectralMatching_delta_sqrt2}}, \figurename{~\ref{fig:SpectralMatching_delta_09}} and \figurename{~\ref{fig:SpectralMatching_delta_07}} show the spectrum of the proposed method with different values of $\delta$. As can be seen in all cases the proposed method is able to perform the spectral masking.

\section{Conclusion} In this paper we discuss about the problem of beampattern shaping with practical constraints in \gls{MIMO} radar systems namely, spectral masking, $3$ dB beam-width, constant modulus and similarity constraints. Solving this problem, not considered hitherto, enables us to control the performance of \gls{MIMO} radar in three domains namely, spatial, spectral and orthogonality (by similarity constraints). Accordingly, we consider a waveform design approach for beampattern shaping optimization problem under. The aforementioned problem leads to a non-convex and NP-hard optimization problem. In order to solve the problem, first by introducing a slack variable we convert the optimization problem to a linear problem with a rank one constraint. Then to tackle the the we proposed an iterative method to obtain the rank one solution. Numerical results shows that the proposed method is able to manage the resources efficiently to obtain the best performance.

% In this paper we discuss about the importance of the resources management in cognitive \gls{MIMO} radar systems. Typically three important resources are spatial, spectrum and time and beampattern shaping, spectral shaping and auto- and cross-correlation minimization are common approaches to manage these three resources respectively. Accordingly, we consider a waveform design approach for beampattern shaping optimization problem under spectral masking and similarity constraints to manage those three types of resources. The aforementioned problem leads to a separable \gls{QCQP}, which is a non-convex and NP-hard optimization problem. In order to solve the problem, first by introducing a slack variable we convert the \gls{QCQP} to a linear problem with a rank one constraint. Then to tackle the the we proposed an iterative method to obtain the rank one solution. Numerical results shows that the proposed method is able to manage the resources to obtain  the best performance.

% or


\appendices
\section{}\label{app:1}
It is readily confirmed that the constraint $\mathbf{X}_n = \bar{\mathbf{s}}_n\bar{\mathbf{s}}_n^{\dagger}$ is equivalent to $\text{Rank} (\mathbf{X}_n - \bar{\mathbf{s}}_n\bar{\mathbf{s}}_n^{\dagger}) = 0$. Further, it can be equivalently expressed as $1 + \text{Rank} (\mathbf{X}_n - \bar{\mathbf{s}}_n\bar{\mathbf{s}}_n^{\dagger}) = 1$. Since $1$ is positive definite, it follows from the Guttman rank additivity formula \cite{gowda2010schur} that $1 + \text{Rank} (\mathbf{X}_n - \bar{\mathbf{s}}_n\bar{\mathbf{s}}_n^{\dagger}) = \text{Rank}(\mathbf{Q}_n)$. Moreover, it follows from $\mathbf{X}_n = \bar{\mathbf{s}}_n\bar{\mathbf{s}}_n^{\dagger}$ and $1 \succ 0$ that $\mathbf{Q}_n$ has to be positive semi-definite. These imply that the constraint $\mathbf{X}_n = \bar{\mathbf{s}}_n\bar{\mathbf{s}}_n^{\dagger}$ in \eqref{eq:P2} can be replaced with a rank and semi-definite constraints on matrix $\mathbf{Q}_n$. Hence, the optimization problem \eqref{eq:P2} can be recast as follows,

\begin{subequations}\label{eq:P4}
  \begin{empheq}[left=\empheqlbrace]{align}
    \label{eq:P4_a} \min_{\mathbf{S}, \mathbf{X}_n} & \quad  \sum_{n=1}^{N}\text{Tr}(\mathbf{A}_u\mathbf{X}_n) \\ 
    \label{eq:P4_b} s.t. & \quad \eqref{eq:P2_b}, \eqref{eq:P2_c}, \eqref{eq:P2_d}, \eqref{eq:P2_e}, \eqref{eq:P2_f}, \eqref{eq:P2_g}\\
    \label{eq:P4_h} & \quad \mathbf{Q}_n \succcurlyeq \ZEROV,\\
    \label{eq:P4_i} & \quad \text{Rank}(\mathbf{Q}_n) = 1,
  \end{empheq}
\end{subequations}

Now, we show that the optimization problem \eqref{eq:P3} is equivalent to \eqref{eq:P4}. Let $\rho_{n,1} \leq \rho_{n,2} \leq \cdots \leq \rho_{n,M+1}$ and $\nu_{n,1} \leq \nu_{n,2} \leq \cdots \leq \nu_{n,M}$ denote the eigenvalues of $\mathbf{Q}_n$ and $\mathbf{V}_n^{\dagger}\mathbf{Q}_n\mathbf{V}_n$, respectively. From the constraint $b_n\mathbf{I}_M - \mathbf{V}_n^{\dagger}\mathbf{Q}_n\mathbf{V}_n \succcurlyeq 0$, we have $\nu_{n,i} \leq b_n, i = 1,2,\cdots, M$ for any $\mathbf{V}_n$ and $\mathbf{Q}_n$ in the feasible set of \eqref{eq:P3}. Additionally, it follows from \cite[Corollary 4.3.16]{horn2012matrix} that $0 \leq \rho_{n,i} \leq \nu_{n,i}, i = 1,2,\cdots, M$ for any $\mathbf{V}_n$ and $\mathbf{Q}_n$ in the feasible set of \eqref{eq:P3}. Hence, we observe that,

\begin{equation}\label{eq:new-CH5}
    \begin{aligned}
    \mathbf{0} &\preccurlyeq \text{Diag}([\rho_{n,1}, \cdots, \rho_{n,M}]^T) \\
    &\preccurlyeq \text{Diag}([\nu_{n,1}, \cdots, \nu_{n,M}]^T) \preccurlyeq b_n \mathbf{I}_M,
    \end{aligned}
\end{equation} 

for any $\mathbf{V}_n$ and $\mathbf{Q}_n$ in the feasible set of \eqref{eq:P3}. It is easily observed from \eqref{eq:P3} and \eqref{eq:new-CH5} that, by properly selecting $\eta$, the optimum value of $\mathbf{V}_n$ will be equal to the eigenvectors of $\mathbf{Q}_n$ corresponding to its $M$ smallest eigenvalues and the optimum values of $b_n, \rho_{n,1}, \cdots, \rho_{n,M}, \nu_{n,1}, \cdots, \nu_{n,M}$ will be equal to zero. This implies that the optimum value of $\mathbf{Q}_n$ in \eqref{eq:new-CH5} possesses one nonzero and $M$ zero eigenvalues. This completes the proof.

% % you can choose not to have a title for an appendix
% % if you want by leaving the argument blank
% \section{}

% % use section* for acknowledgment
% \section*{Acknowledgment}


\ifCLASSOPTIONcaptionsoff
  \newpage
\fi



%\bibliographystyle{IEEEtran}


% \end{thebibliography}

\bibliographystyle{IEEEtran}
\bibliography{IEEEabrv,Main_Arxiv}


% \end{IEEEbiography}


%\newpage



%\vfill


\end{document}